\documentclass[iop]{emulateapj}

\usepackage{amsmath,amssymb,latexsym,graphics,epsfig,exscale,graphicx}
\usepackage{pstricks} \usepackage{multirow} \usepackage{graphicx}
\usepackage{subfigure}

\newcommand{\gad}{{\sc Gadget-2}} 
\newcommand{\hmpc}{h^{-1}{\rm Mpc}} \newcommand{\lcdm}{$\Lambda$CDM}
\newcommand{\Msun}{$M_{\odot}$} 
\newcommand{\Mbulge}{$M_{\rm BH}$--$M_{\rm bulge}$}

\newcommand{\Ngal}{213}

\newcommand{\pathI}{figures}

\begin{document}

\title{Torque-limited growth of massive black holes in galaxies across cosmic time} 
\shorttitle{Torque-Limited Growth of Black Holes in Galaxies}

\shortauthors{D. Angl{\'e}s-Alc{\'a}zar et al.} \author{Daniel
Angl{\'e}s-Alc{\'a}zar\altaffilmark{1,2},
            Feryal {\"O}zel\altaffilmark{3}, Romeel
            Dav{\'e}\altaffilmark{4,5,6}, Neal Katz\altaffilmark{7},
            Juna A. Kollmeier\altaffilmark{8}, and Benjamin
            D. Oppenheimer\altaffilmark{9}}
\altaffiltext{1}{Department of Physics, University of Arizona, Tucson, AZ 85721, USA; anglesd@email.arizona.edu}
\altaffiltext{2}{Center for Interdisciplinary Exploration and Research in Astrophysics (CIERA) and Department of Physics and Astronomy, Northwestern University, Evanston, IL 60208, USA}
\altaffiltext{3}{Astronomy Department, University of Arizona, Tucson, AZ 85721, USA} 
\altaffiltext{4}{University of the Western Cape, Bellville, Cape Town 7535, South Africa} 
\altaffiltext{5}{South African Astronomical Observatories, Observatory, Cape Town 7925, South Africa} 
\altaffiltext{6}{African Institute for Mathematical Sciences, Muizenberg, Cape Town 7945, South Africa} 
\altaffiltext{7}{Astronomy Department, University of Massachusetts, Amherst, MA 01003, USA} 
\altaffiltext{8}{Observatories of the Carnegie Institute of Washington, Pasadena, CA 91101, USA} 
\altaffiltext{9}{CASA, Department of Astrophysical and Planetary Sciences, University of Colorado, Boulder, CO 80309, USA}

\begin{abstract}

We combine cosmological hydrodynamic simulations with analytic models to
evaluate the role of galaxy-scale gravitational
torques on the evolution of massive black holes at the centers of
star-forming galaxies.  We confirm and extend our earlier results to show
that torque-limited growth yields black holes and host galaxies evolving
on average along the \Mbulge~relation from early times down to $z = 0$ and
that convergence onto the scaling relation occurs independent of the initial
conditions and with no need for mass averaging through mergers or additional
self-regulation processes.  Smooth accretion dominates the long-term evolution,
with black hole mergers with mass ratios $\gtrsim$1:5 representing typically a
small fraction of the total growth.  Winds from the accretion disk are required
to eject significant mass to suppress black hole growth, but there is no need
for coupling this wind to galactic-scale gas to regulate black holes in a
nonlinear feedback loop.  
Torque-limited growth yields a close-to-linear $\langle \dot{M}_{\rm BH}
\rangle \propto$ star formation rate (SFR) relation for the black hole accretion rate averaged
over galaxy evolution timescales.  However, the SFR--AGN connection has
significant scatter owing to strong variability of black hole accretion at all
resolved timescales.  Eddington ratios can be described by a broad lognormal
distribution with median value evolving roughly as $\lambda_{\rm MS} \propto
(1+z)^{1.9}$, suggesting a main sequence for black hole growth similar to the
cosmic evolution of specific SFRs.  Our results offer an attractive scenario
consistent with available observations in which cosmological gas infall and
transport of angular momentum in the galaxy by gravitational instabilities
regulate the long-term co-evolution of black holes and star-forming galaxies.

\end{abstract}


%%%%%%%%%%%%%%%%%%%%%%%%%%%%%%%%%%%%%%%%%%%%%%%%%%%%%%%%%%%%%%%%%%%%%%%%%%%%%%%%%%%%%%%%%%%%%%%%%%%%%%%%%%%%%%%%%%%%%%%%%%%%%%%%

\section{Introduction}

\setcounter{footnote}{0}

A wide range of observations imply a close connection between central massive
black holes and their host galaxies, including the similarity between
the cosmic star formation history and the evolution of global black hole
accretion \citep{madau96,boy98,hopk06,silverman08,aird10,rodighiero10},
the higher incidence of active galactic nuclei (AGNs)
in higher-mass galaxies and strongly star-forming systems
\citep{kauffmann03,silverman09,raff11,santini12,juneau13,rosario13,trump13},
as well as a number of correlations between the mass of the central black
hole and properties of the host galaxy such as the stellar mass of the
central bulge \citep[\Mbulge~relation;][]{magorrian98,har04,scott13}, and its
velocity dispersion \citep{ferrarese00,tremaine02,gultekin09,gra11,macconnell11}.
This circumstantial evidence has led many to conclude that massive black holes
play a key role in galaxy evolution
\citep{somerville08,cattaneo09} and yet, unravelling the physical mechanisms
driving this connection remains one of the major unsolved problems in modern
astrophysics.

The growth of massive black holes at the centers of galaxies involves a
remarkable variety of physical processes operating at scales ranging from the
size of the entire galaxy down to the black hole event horizon \citep[see][for
a review]{alexander12}.  In a broad view, the procedure for growing black
holes involves (1) feeding the black hole from the accretion disk;
(2) the regulation of growth owing to feedback processes (winds and
thermal pressure); and (3) the supply of gas from the galaxy onto
the accretion disk.

An accretion flow forms in the region where the potential of the black
hole dominates that of the galaxy and angular momentum is transported
outward by turbulent MHD processes \citep{shakura73,balbus98}.
The rate at which gas inflows through the sphere of influence of the
black hole is believed to determine the overall geometry and radiative
properties of the accretion flow (see, e.g., \citealt{abramowicz13}
for a recent review). In different accretion rate regimes, analytic
arguments and numerical simulations show that a significant fraction of
the inflowing mass is likely to be lost to winds and outflows \citep[see,
e.g.,][]{blandford82,narayan95a,proga00,narayanan06,ohsuga11,sadowski13}.
Moreover, observations show that winds and outflows are frequent in AGNs
\citep{reynolds97,veilleux05,fabian12} and may carry significant amounts of
mass away \citep[e.g.,][]{king13}.  Indeed, powerful galactic-scale molecular
gas outflows thought to be driven by nuclear activity are observed both in the
local and the high redshift universe, and the total mass loss rate may exceed
the star formation rate (SFR) of the entire galaxy \citep{feruglio10,rupke11,sturm11,maiolino12}.
Thus, winds and outflows powered by black hole accretion could represent a
significant mass loss relative to the inflowing gas from larger scales.

The impact of this outflowing gas on the scale of the galaxy has received
considerable attention recently, as a way to further regulate black hole growth
by actively affecting the rate at which gas inflows feed the accretion disk
from galactic scales.  If the interaction between inflows and the outflowing
mass and radiation is strong enough, this ``AGN feedback" may be the primary
modulator of long term black hole growth \citep[e.g.,][]{fabian12}.  In this
scenario, black hole growth becomes self-regulated, the feedback coupling
efficiency represents the key physical process, and the mechanism responsible
for driving gas inflows from galactic scales down to the accretion flow
becomes sub-dominant.  This paradigm has been extensively explored both
in analytic models \citep[e.g.,][]{silk98,king03,murray05} and numerical
simulations \citep{dimatteo05,hop06,robertson06b,hop07,dimatteo08,booth09,dubois12}
with significant success in explaining a variety of observations including
the black hole--galaxy scaling relations.

AGN feedback may also have a strong impact on the host galaxy and
possibly be responsible for the observed exponential cutoff at the high
mass end of the stellar mass function \citep{baldry08} and the observed
dichotomy between blue star-forming galaxies and red quiescent galaxies
\citep{schawinski07}.  Indeed, AGN feedback is often invoked in semianalytic
models \citep[e.g.,][]{bower06,croton06,somerville08} and hydrodynamic
simulations \citep[e.g.,][]{spr05b,gabor11,teyssier11,dubois13,puchwein13}
as an additional energy source to suppress cooling flows and star formation in
early-type galaxies.  There remain, however, significant concerns relative
to the overall efficiency of feedback required by self-regulated models
\citep[e.g.,][]{silk10}, the interplay between AGN feedback and stellar
feedback \citep[e.g.,][]{cen12}, and the intrinsic degeneracy often suffered
by coupled accretion-feedback models \citep{newton13,wurster13}.

In comparison, the physical processes responsible for feeding the
black hole accretion disk in the first place have received comparably
little attention.  Most numerical investigations have relied on the
Bondi--Hoyle--Littleton accretion prescription~\citep{hoy39,bon44,bon52} to
capture gas from the inner galaxy and feed the black hole accretion disk
\citep[e.g.,][]{dimatteo05,booth09}.  However, this prescription does not
account for the rate at which angular momentum can be lost by the infalling
gas, which could easily be the limiting factor for fuelling AGNs \citep{jogee06}.
Hence, the physical mechanisms driving the required continuous supply of gas
from galactic scales down to sub-parsec scales may play a more crucial role
than commonly considered \citep{escala06,escala07}.

Hydrodynamic simulations of gas-rich galaxy mergers have shown that
large-scale tidal torques induced by the interaction, or even gravitational
instabilities in self-gravitating disks, can lead to angular momentum
transport and the rapid inflow of gas to the central $\sim 100$\,pc of
galaxies \citep{hernquist89,shlosman89,barnes92,escala07,hop10}.  Another
alternative for AGN fuelling is direct clump--clump interactions in turbulent
gas-rich disks at high redshift \citep{bournaud11,gabor13}.  However,
subsequent gravitational instabilities become less efficient at scales
comparable to the black hole radius of influence, $\sim 10$\,pc, requiring
additional mechanisms to transport gas down to smaller scales \citep{jogee06}.
Furthermore, the gas is still self-gravitating at these scales and, therefore,
likely to participate in star formation \citep{thompson05}.  Using multiple
nested simulations of progressively higher resolution, \citet{hop10,hop11}
showed that non-axisymmetric perturbations to the stellar potential may
induce strong orbit crossing, driving gas into shocks that dissipate energy
and angular momentum, and providing significant gas inflows down to $\sim
0.01$\,pc scales.

In this paper, we evaluate the role of black hole feeding limited by
galaxy-scale gravitational torques on the evolution of massive black holes at
the centers of star-forming galaxies over cosmic time,
minimizing the assumptions made on the effects of AGN feedback on galactic scales.  
In our previous work
\citep{ang13}, we combined cosmological zoom simulations of galaxy formation
down to $z = 2$ together with analytic parametrizations of black hole growth
to show that a model in which black hole growth is limited by galaxy-scale
torques \citep{hop11} does not require self-regulation of black hole growth.
Specifically, torque-limited growth yields black holes and galaxies evolving
on average along the observed scaling relations from early times down to $z
\sim 2$, providing a plausible scenario to explain their connection that
does not crucially invoke AGN feedback.  Winds from the accretion disk
are still required in this scenario to drive significant mass
loss from the accretion disk (roughly 95\%), thereby strongly suppressing
black hole growth, but there is no need to strongly couple these winds to
galaxy-scale gas to regulate black hole growth in a nonlinear
feedback loop.  
This removes the need for self-regulation via spherical feedback as commonly assumed in
Bondi accretion-based models\footnote{See \citet{dubois12} for a non-isotropic kinetic-mode feedback model capable of self-regulating black hole growth in the context of Bondi accretion.}; 
instead, the wind can propagate biconically from the accretion disk and
be weakly coupled to the inflow at a sub-resolution level, which is perhaps more physically plausible
for black hole growth within disk galaxies.

Motivated by the attractive features of the torque-limited growth model,
we extend the analysis in \citet{ang13} to examine black hole growth in a
larger population of galaxies down to $z=0$ by employing full cosmological
hydrodynamic simulations.  We describe the simulations and the overall
methodology in Section~\ref{sec:methods} and report our main results in
Section~\ref{sec:results}.  We present resolution convergence tests to show
the robustness of our methodology in Section~\ref{sec:res}, and we conclude
in Section~\ref{sec:dis} by discussing implications in the context of current
theoretical models and observations.

%%%%%%%%%%%%%%%%%%%%%%%%%%%%%%%%%%%%%%%%%%%%%%%%%%%%%%%%%%%%%%%%%%%%%%%%%%%%%%%%%%%%%%%%%%%%%%%%%%%%

\section{Methodology}\label{sec:methods}

We apply and extend the methodology described in \citet{ang13} 
to follow the growth of massive black holes over cosmic time.  We begin by
identifying a population of galaxies at $z = 0$ from a full cosmological
hydrodynamic simulation and characterize their evolution back in time.  Then,
we infer how black holes grow at the centers of galaxies in post-processing,
by evaluating accretion rates based on the gravitational torque model of
\citet{hop11}, and accounting for the mass growth through black hole mergers.

\subsection{Simulations}

We use an extended version of the $N$-body + smoothed particle hydrodynamics
cosmological galaxy formation code \gad~\citep{spr05} to simulate the
evolution of a $[32\,\hmpc]^3$ comoving volume down to $z = 0$.  Our primary
simulation utilizes $2 \times 512^3$ gas + dark matter particles with masses
$m_{\rm gas} \approx 4.5 \times 10^6$\,\Msun~and $m_{\rm DM} \approx 2.3
\times 10^7$\,\Msun, respectively, and a fixed comoving softening length
$\epsilon \approx 1.25\,h^{-1}$\,kpc.  Throughout this paper we assume a
\lcdm~concordance cosmology with parameters $\Omega_{\rm \Lambda} = 0.72$,
$\Omega_{\rm M} = 0.28$, $\Omega_{\rm b} = 0.046$, $h = 0.7$, $\sigma_{8}
= 0.82$, and $n = 0.96$, consistent with the latest nine-year {\it Wilkinson Microwave Anisotropy Probe} data
\citep{hin13}.

Our main simulation has been first described in \citet{dav13}.  We include
radiative cooling from primordial gas \citep{kat96}, metal-line cooling
\citep{sut93}, and photoionization heating from an optically thin UV
background \citep{haa01} starting at $z = 9$.  Star formation is modeled
probabilistically through a multi-phase sub-grid prescription \citep{spr03}
where gas particles that are sufficiently dense to become Jeans unstable can
spawn a star particle with a probability based on a \citet{schmidt59} law.
The resulting SFRs are tuned to be in accord with the observed \citet{ken98}
relation.  We include metal enrichment from Type Ia and Type II supernovae (SNe) and
asymptotic giant branch (AGB) stars,  energy feedback from Type Ia and Type
II SNe, and mass-loss from AGB stars as described in \citet{opp06,opp08}.
We assume a \citet{cha03} initial mass function throughout.

Galactic outflows are modeled by imparting kinetic energy to gas particles
with a probability given by the mass loading factor ($\eta$) times the star
formation probability.  Outflow velocities scale with galactic velocity
dispersion ($\sigma$) and the mass loading factor scales as $\eta \propto
1/\sigma$ (as in the momentum-driven case) and $\eta \propto 1/\sigma^2$
(as in the energy-driven case) for galaxies above and below $\sigma =
75$\,km\,s$^{-1}$, respectively \citep{dav13}.  This is motivated by recent
analytic models \citep{murray10} as well as galaxy-scale hydrodynamic
simulations with explicit stellar feedback models \citep{hop12}.  Our primary
simulation also incorporates a heuristic prescription to quench star
formation that is tuned to reproduce the observed exponential cutoff in
the high-mass end of the stellar mass function at $z = 0$ \citep{dav13}.
This ad hoc quenching prescription has no major effect on our results,
as we show in Section~\ref{sec:res}.

Note that we do not attempt to explicitly model AGN feedback in our
simulations.  Instead, we focus on the role of feeding black holes by
galaxy-scale gravitational torques and use the observed connections between
central black holes and host galaxies to put constraints on the overall
impact of AGN feedback.

\subsection{Host Galaxies}

We produce 135 redshift snapshots from $z = 30$ down to $z = 0$.
Following \citet{ang13}, we identify individual galaxies in each
snapshot as bound collections of star-forming gas and star particles
by means of the Spline Kernel Interpolative Denmax algorithm ({\sc
skid}\footnote[1]{http://www-hpcc.astro.washington.edu/tools/skid.html}).
Each {\sc skid}-identified galaxy is associated with a dark matter halo by
using a spherical overdensity algorithm, where the virial radius is defined
to enclose a mean density given by \citet{kitayama96}.  Overlapping halos
are merged together so that every final halo contains one central galaxy
(the most massive galaxy) and a number of satellite galaxies by construction.

We follow the evolution of central galaxies back in time beginning at $z =
0$ by identifying their most massive progenitor at each previous snapshot.
The main progenitor at time $t$ is defined as the galaxy with the highest
fraction of the total stellar mass of a given galaxy at time $t + \Delta t$.
Only a sub-sample of all central galaxies identified at $z = 0$ is used in our
primary analysis.  Unless otherwise noted, we require galaxies to contain at
least 200 gas and 200 star particles at all times and to be identified in the
cosmological simulation as early as $z \geq 4$.  This selection criteria allows
us to characterize the morphological properties of galaxies and to evaluate
the evolution of their central black holes for a cosmologically significant
period of time.  Nonetheless, in Sections~\ref{sec:sfagn} and~\ref{sec:res}
we will enlarge our galaxy sample to expand the dynamic range for
a few particular redshifts.

\begin{figure} 
\begin{center}
\includegraphics[scale=1]{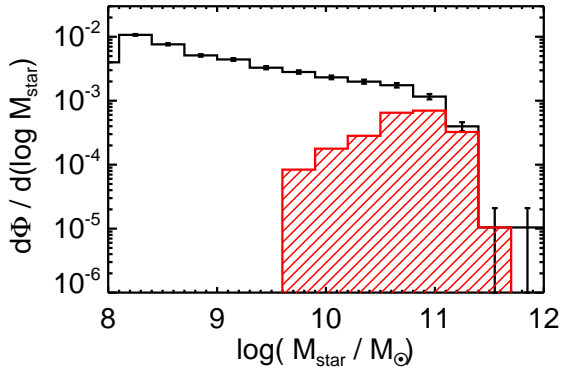} 
\end{center} 
\caption{Galaxy stellar mass function at $z = 0$ (black).  The red hatched area corresponds to the
primary galaxy sample used in this work.} 
\label{fig:massfunc} 
\end{figure}

Figure~\ref{fig:massfunc} shows the stellar mass function for all galaxies in
our $[32\,\hmpc]^3$ simulated volume, where the red hatched area indicates the
primary sub-sample of \Ngal~galaxies selected for this work.  As expected,
the requirement for galaxies to be resolved in the simulation at $z \geq 4$
results in a sub-sample containing mainly massive galaxies.  
Note that the
requirement for a minimum number of gas particles eliminates eleven massive
galaxies with extremely low gas fractions at low redshift.

\subsection{Black Hole Seeds}\label{sec:ini}

Several alternative scenarios have been proposed for the formation of
primordial seeds that eventually become the massive black holes populating the
centers of galaxies \citep[for a review, see][]{volonteri10}.  Popular models
include the formation of light seeds ($\sim 10^{2}$\,\Msun) as remnants of
population III stars \citep[e.g.,][]{madau01} and the formation of massive
black holes ($\sim 10^{5}$\,\Msun) by direct collapse in pre-galactic halos
\citep[e.g.,][]{begelman06,choi13}.  Despite much theoretical work, major
uncertainties remain on the initial mass of black hole seeds, their birth
places and number densities, and their formation redshift.

A common feature of current theoretical models is the requirement of large
amounts of pristine gas only available at very high redshifts ($z \gtrsim 15$).
Since our simulations do not resolve galaxies until $z \approx 8$ even for
the most massive systems, we have to populate galaxies with black holes that
have been presumably evolving within their hosts for at least a few hundred
million years.  The simplest approach is, therefore, to assume that there is
only one central black hole by the time we first resolve each galaxy and that
its mass scales with the stellar mass of the host galaxy in a way similar to
the observed $z = 0$ \Mbulge~relation \citep[dual AGN are usually associated
with merger systems, at least at low redshift; see, e.g.,][]{comerford09}.

We assign a seed black hole to every galaxy by assuming consistency with the
\Mbulge~relation of \citet{har04} evaluated for the stellar mass within the
effective radius of the host galaxy, regardless of the redshift when it is
first resolved in the simulation.  We also perform tests by assigning initial
black hole masses either a factor of 10 above or below the scaling relation,
or drawn from a log-normal distribution with a mean and a dispersion similar to
that of the observed local \Mbulge~relation.  We will justify the assumption
of correlated initial conditions in the torque-limited growth model, since
we will show that it yields black holes that evolve toward the observed
scaling relations independent of their initial conditions \citep{ang13}.

\subsection{Accretion Rates}\label{sec:accr}

Once a seed black hole has been assigned to a given galaxy, accretion rates are
calculated based on the gravitational torque rate introduced by \citet{hop11},
$\dot{M}_{\rm Torque}$, assuming that only a fraction $\epsilon_{\rm m}$
of the inflowing gas at sub-parsec scales is actually accreted by the black
hole, with the rest lost to winds and outflows \citep{ang13}:

\begin{equation}\label{eq:corr} 
\frac{dM_{\rm BH}}{dt} = \epsilon_{\rm m} \, \dot{M}_{\rm Torque}(t) 
\end{equation}

The gravitational torque model predicts gas inflow rates from galactic
scales to sub-parsec scales as a function of galaxy properties evaluated
within a radial aperture, $R_{0}$, that must be resolved in the cosmological
simulation \citep{hop11}:

\begin{align}\label{eq:torque} & \dot{M}_{\rm Torque} \approx \alpha_{\rm T} \,
f_{\rm d}^{5/2} \times \left ( \frac{M_{\rm BH}}{10^{8}\,{\rm M_{\odot}}}
\right )^{1/6} \left ( \frac{M_{\rm d}(R_{0})}{10^{9}\,{\rm M_{\odot}}}
\right ) \nonumber \\
 & \times \left ( \frac{R_{0}}{100\,{\rm pc}} \right )^{-3/2}  \left (1 +
 \frac{f_{0}}{f_{\rm gas}} \right )^{-1} \, {\rm M_{\odot}\,yr^{-1}},
\end{align}  
where 
\begin{equation} 
f_{\rm d} \equiv M_{\rm d}(R_{0}) / (M_{\rm gas}(R_{0})+M_{\rm star}(R_{0})), 
\end{equation}
\begin{equation}\label{eq:fgas}
f_{\rm gas} \equiv M_{\rm gas}(R_{0})/M_{\rm d}(R_{0}),
\end{equation} 
\begin{equation} f_{0} \approx 0.31 \, f_{\rm d}^{2} \, (M_{\rm d}(R_{0})/10^{9}{\rm M_{\odot}})^{-1/3}, 
\end{equation}
and $M_{\rm d}(R_{0})$ is the total (gas+stars) disk mass within $R_{0}$,
$M_{\rm gas}(R_{0})$ and $M_{\rm star}(R_{0})$ represent the total gas
and stellar masses within $R_{0}$, and $\alpha_{\rm T} \approx 5$ is a
normalization factor that parametrizes the dependence of inflow rates on
star formation at scales not resolved \citep{hop11,ang13}.

To estimate the disk mass within $R_{0}$ for the gas and stellar components, $M_{\rm d}$, we perform a simple bulge--disk kinematic decomposition using the full three-dimensional information available in the simulations.  
Recent morphological studies of simulated galaxies have identified two distinct dynamical components in the distribution of the rotational support of their baryonic content, clearly associated with the disk and bulge morphological components \citep[e.g.,][]{abadi03,governato09,hop09,scannapieco09,christensen14}.
Motivated by these studies, we calculate the azimuthal velocity $v_{\rm \phi}$ of each particle
with respect to the direction of the total angular momentum within $R_{0}$,
and estimate the mass in a spheroidal component, $M_{\rm bulge}(R_{0})$,
as double the mass of particles moving with $v_{\rm \phi} < 0$.  The disk
mass is, then, $M_{\rm d}(R_{0}) \approx M_{\rm tot}(R_{0}) - M_{\rm
bulge}(R_{0})$, where $M_{\rm tot}(R_{0}) = M_{\rm gas}(R_{0}) + M_{\rm
star}(R_{0})$ is the total mass within $R_{0}$.
Note that this kinematic decomposition is formally equivalent to that performed in \citet{abadi03} based on the distribution of the orbital circularity parameter.
The basic assumption is that the spheroid has little net rotation, with as many gas/star particles in co- as in counterrotating orbits.
This will certainly overestimate $f_{\rm d}$ in the case of rotating bulges but it is a reasonable approximation for the purpose of evaluating Equation~(\ref{eq:torque}).
While several different bulge--disk decomposition procedures are possible, our main results are qualitatively independent of the exact definition of the bulge and disk components.  Any quantitative differences could be in principle absorbed into the normalization factor $\epsilon_{\rm m}$, and, as we show in the Appendix, our bulge--disk decomposition procedure shows better resolution convergence relative to other methods.

\begin{figure*} 
\begin{center}
\includegraphics[scale=0.9]{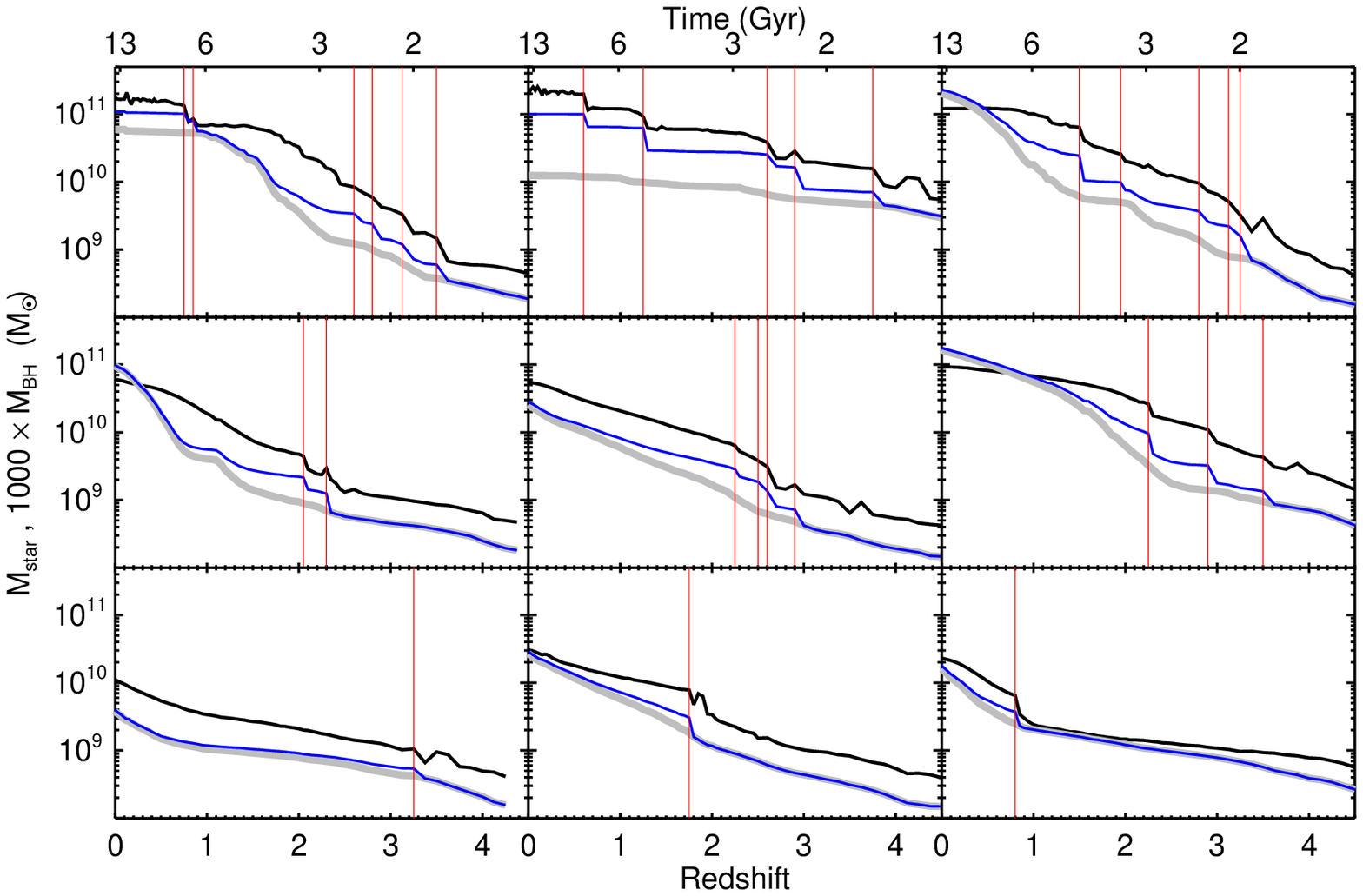} 
\end{center} 
\caption{Evolution of the total stellar mass (black lines) and the central black hole
mass (blue and gray lines; see below) for nine representative galaxies. 
Black holes grow according to the
gravitational torque rate (Equation~(\ref{eq:corr})) with a mass retention
rate $\epsilon_{\rm m} = 0.05$.  Initial black hole seeds are taken to be
consistent with the \Mbulge~relation \citep{har04} evaluated for the stellar
mass within the effective radius of the host galaxy.  Black hole mergers are
assumed to occur after major galaxy mergers with stellar mass ratios above 1:5,
which are indicated by the red vertical lines.  Gray lines correspond to black
hole growth from torque-limited accretion only (upscaled by a factor of 1000),
while blue lines show the total black hole growth including mergers
(also upscaled by 1000), where we assume that the merging
galaxy has a central black hole with a mass consistent with the corresponding
\Mbulge~relation.} 
\label{fig:bhm} 
\end{figure*}

In our previous work, we found that a constant radial aperture
$R_{0} = 1$\,kpc to be
appropriate for all galaxies at all times, since kiloparsec scales were well
resolved in our cosmological zoom simulations \citep{ang13,ang14}.  This
fixed radial aperture is likely not appropriate here given the significantly
larger range of galaxy masses and evolution times.
Instead, we adopt a variable, time-dependent $R_{0}$ defined to be the smallest
radial aperture containing at least 200 gas particles and 200 star particles.
With this definition, we ensure that physical quantities such as gas fraction
and disk fraction entering into the calculation of gravitational torque
rates can be appropriately characterized for all galaxies at all times.
In Section~\ref{sec:res}, we evaluate the effects of using different radial
apertures on the inferred black hole accretion rates, and show that it has
only a modest impact.

We calculate the growth of black holes through direct smooth gas accretion
by numerical integration of Equation~(\ref{eq:corr}) for the initial black
hole mass defined for each galaxy (Section~\ref{sec:ini}).  The integration
time step is constrained by the number of redshift snapshots available,
ranging in frequency from $\sim 10$ to 300\,Myr in the redshift range $z \sim
6 \rightarrow 0$, i.e. $\la 2$\% of the Hubble time at any given redshift.
Inferred black hole accretion rates represent, therefore, average values
for the corresponding time steps.  The gravitational torque rate, $\dot{M}_{\rm Torque}(t)$, is calculated based on the physical properties of each galaxy
at a given time (Equation~(\ref{eq:torque})) and is evaluated with the
appropriate black hole mass at each time step, as given by Equation~(\ref{eq:corr}).

Note that by evaluating Equation~(\ref{eq:corr}) in post-processing we are neglecting the gravitational influence of the central black hole at the scales resolved in the simulation.  This is unlikely to affect our results since we are considering the transport of angular momentum at scales well beyond the black hole radius of influence.  
In addition, we assume that outflows powered by black hole accretion are weakly coupled to the gas inflows and do not alter significantly the evaluation of gravitational torque rates \citep{ang13}.
Equation~(\ref{eq:corr}) implies that a total mass $M_{\rm out} \approx (1/\epsilon_{\rm m}-1) \times M_{\rm BH}$ will be ejected from the accretion disk during the full evolution of the central black hole, though not necessarily leaving the host galaxy.  This represents $\sim 2$\,\% of the final stellar mass of the host galaxy, simply assuming $M_{\rm BH} \sim M_{\rm star} / 1000$ and $\epsilon_{\rm m} = 5\,\%$ (see Section~\ref{sec:msig}).  In contrast, for a mass loading factor $\eta \sim 2$, star formation driven winds in our simulation will have ejected at least 100 times more gas than the direct mass loss owing to accretion-driven winds (assuming no significant entrainment of cold ISM gas).  It is, thus, reasonable to treat the overall mass loss $\epsilon_{\rm m}$ as a zeroth order effect on black hole growth and neglect any higher order effects for the purpose of evaluating $\dot{M}_{\rm Torque}$.

\subsection{Black Hole Mergers}\label{sec:merg}

In addition to black hole growth by torque-limited accretion, we evaluate the
mass growth rate from black hole mergers.  If a major merger is identified
for a given central galaxy, we assume that the merging galaxy contains a
black hole consistent with the \Mbulge~relation of \citet{har04} and we
add its corresponding mass to the total mass of the final black hole in the
remnant galaxy.  We also incorporated an alternate prescription as with black
hole seeding, where the mass of the merging black hole is chosen randomly
from a log-normal distribution corresponding to the \Mbulge~relation for
its host galaxy and added that mass to the central black hole accordingly.
Any time delay between the merger of the host galaxies and the final merger of their central black holes \citep[e.g.,][]{dubois10} is neglected for the sake of simplicity.  This is unlikely to affect our results given the weak dependence of gravitational torque rates on black hole mass.

We limit ourselves to major galaxy mergers where the mass ratios of interacting
galaxies are above 1:5.  If $M_{*}(t+\Delta t)$ is the total stellar mass of
a central galaxy at time $t+\Delta t$ (where $\Delta t$ represents the time
interval between simulations outputs) and $M_{*}^{1st}(t)$ and $M_{*}^{2nd}(t)$
are the stellar masses of its first and second most massive progenitors at time
$t$, major galaxy mergers ($>$1:5) are identified by the following criteria:

\begin{figure*} 
\begin{center}
\includegraphics[scale=0.8]{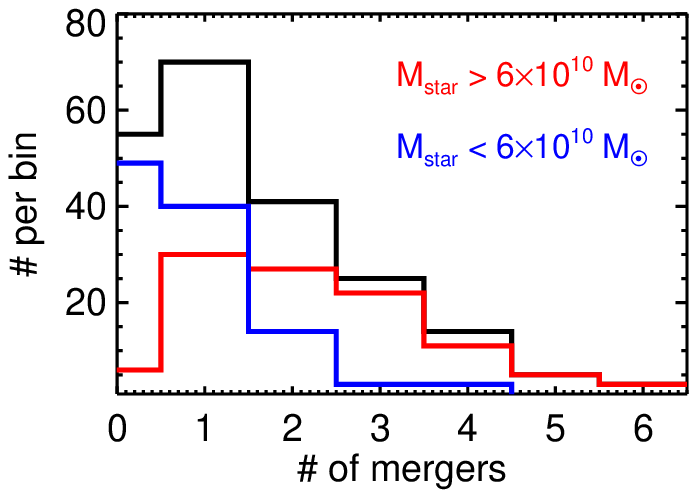}
\includegraphics[scale=0.8]{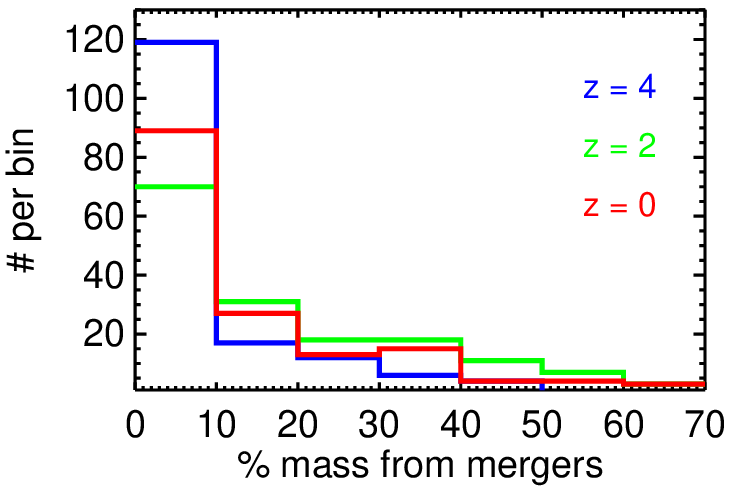} 
\end{center} 
\caption{{\it Left}: distribution of galaxies in terms of the number of major mergers
(with stellar mass ratios $>$1:5) down to $z = 0$, for the
full galaxy sample (black), for galaxies with stellar masses $M_{\rm star} >
6\times10^{10}$\,\Msun~(red), and for galaxies with stellar masses
$M_{\rm star}
< 6\times10^{10}$\,\Msun~(blue).  {\it Right}: distribution of black holes
undergoing one or more mergers relative to the percentage of the 
mass contributed
by black hole mergers, computed at $z = 0$ (red), $z = 2$ (green), and
$z = 4$ (blue).   Black hole seeds and merging black holes are assumed to
lie on the \Mbulge~relation for the corresponding host galaxy.  Most black
holes have a mass contribution of less than 10\,\% from black hole mergers.}
\label{fig:hist_mrg} 
\end{figure*}

\begin{enumerate}

\item $M_{*}^{2nd}(t) \geq 1/5 \times M_{*}^{1st}(t)$

\item $M_{*}(t+\Delta t) > (1 + 1/5)\,M_{*}^{1st}(t)$

\item $M_{*}(t+\Delta t) > 0.8\,(M_{*}^{1st}(t) + M_{*}^{2nd}(t))$

\item $min\{\Delta M_{*}\}_{t\to t+600\,{\rm Myr}} > -0.5\,M_{*}^{2nd}(t)$

\end{enumerate}

The identification of galaxy mergers in cosmological simulations is not
a trivial task, where the simple working definition of ``galaxy" can, for
example, result in the wrong identification of close galaxy encounters as
a merging system \citep[e.g.,][]{gabor11}.  We determined and tested the
above criteria experimentally by comparing the identified merger events
against the evolution of the stellar mass of central galaxies relative to
that of their most massive progenitors to ensure that close encounters are
not treated as mergers.

The first and second conditions reflect our definition of major mergers
($>$1:5) and the requirement that the central galaxy has indeed grown by
at least one fifth relative to its stellar mass in the previous time step.
Note that the mass increase $\Delta M_{*} = M_{*}(t+\Delta t) - M_{*}(t)$,
where $M_{*}(t) \equiv M_{*}^{1st}(t)$, contains contributions from both
major and minor mergers as well as star formation within the galaxy.
The third condition, requiring that the merger remnant contains at least
80\,\% of the mass of its two most massive progenitors, is apparently less
restrictive than the second condition; however, it accounts for situations
in which $M_{*}^{2nd}(t) > M_{*}^{1st}(t)$ that may occur if only a small
fraction of $M_{*}^{2nd}(t)$ ends up in the merger remnant.  Finally, the
fourth condition attempts to correct for wrong identifications during close
galaxy encounters by requiring that any decrease in stellar mass during the
$\sim 600$\,Myr after the merger cannot be higher than half the mass
of the second most massive progenitor.

At all times, gravitational torque rates (Equation~(\ref{eq:torque}))
are evaluated according to the current mass of the black hole including
contributions from mergers.  Note that we neglect the possibility of black
holes leaving the center of their host galaxies owing to gravitational recoils
\citep[e.g.,][]{blecha08}.

%%%%%%%%%%%%%%%%%%%%%%%%%%%%%%%%%%%%%%%%%%%%%%%%%%%%%%%%%%%%%%%%%%%%%%%%%%%%%%%%%%%%%%%%%%%%%%%%%%%%%

\section{Results}\label{sec:results}

\subsection{Black Hole Mergers versus Smooth Accretion}

Figure~\ref{fig:bhm} illustrates the identification of galaxy mergers, based
on the criteria described in Section~\ref{sec:merg}, by showing the evolution
of the total stellar mass of nine representative galaxies in the mass range
$M_{\rm star} = 10^{10}$--$2\times 10^{11}$\,\Msun~at $z = 0$.  Major merger
events are indicated by red vertical lines and correspond to abrupt changes
in the stellar mass of the galaxies.  Note that the time interval between
data snapshots varies with redshift, implying that the mass increase per unit
time required for merger identification is redshift dependent.  This could
result in an increasing number of merger identifications at lower redshifts
owing, for example, to contributions from smooth accretion and minor mergers
occurring in a single (longer) time step.  
However, this is compensated by
the shorter time steps at the epoch near the peak of cosmic star formation activity ($z \sim 2$).

Galaxy misidentifications by {\sc skid} represent a more challenging issue
\citep[e.g.,][]{gabor11}.  Interacting galaxies are sometimes identified as
one single galaxy at the closest approach during the first orbital passage,
with a consequent increase in the stellar mass of the newly identified
central galaxy.  When the distance between the interacting galaxies increases
again, two separate systems are identified and the mass of the central galaxy
decreases correspondingly.  The fourth condition for merger identification
in Section~\ref{sec:merg} attempts to correct for this effect.  We present
examples of this for several galaxies in Figure~\ref{fig:bhm}.

Overall, our simple method provides a robust identification of galaxy mergers
and allows us to estimate the contribution of black hole mergers to total
black hole growth.  For each galaxy in Figure~\ref{fig:bhm}, the total mass
of the central black hole as a function of redshift is shown as the blue line
(upscaled by a factor of 1000), while the gray lines correspond to
black hole growth from torque-limited accretion only.  Here we adopt a mass
retention rate $\epsilon_{\rm m} = 0.05$, which has been shown to reproduce
the normalization of the \Mbulge~relation at $z \ge 2$ \citep{ang13},
and assume that the merging galaxy contains a black hole consistent with
the local \Mbulge~relation of \citet{har04}.  We relax this assumption in
Section~\ref{sec:msig} by considering a 0.5 dex scatter in black hole mass.

The most massive black holes are expected to undergo more frequent mergers
as their host galaxies also represent the high-mass end of the galaxy mass
distribution and live in higher density environments.  Indeed, the left
panel of Figure~\ref{fig:hist_mrg} shows that our sub-sample of lower mass
galaxies clearly dominates the population of galaxies that undergo only
one or no major mergers during their entire evolution down to $z = 0$.
Correspondingly, galaxies in the higher-mass sub-sample tend to undergo two
or more major mergers down to $z = 0$.

The contribution from each black hole merger represents a significant
fraction of the total black hole mass at the time of the merger event,
typically $\gtrsim 20\,\%$ given that we define major galaxy mergers to be mass
ratios above 1:5.  Despite this, the continuous supply of gas through smooth
accretion by gravitational torques tends to erase the merger histories of
black holes.  By $z = 0$,  black hole growth from mergers is typically a
small fraction of the total growth, except in some exceptional cases with
numerous mergers happening preferentially at low redshift for which the final
black hole mass may exceed the total accreted mass by factors of a few (e.g.
see the top left and the top middle panels of Figure~\ref{fig:bhm}).

This is quantified more rigorously in the right panel of
Figure~\ref{fig:hist_mrg}, where we show the distribution of black holes in
terms of the percentage of mass contributed by mergers, evaluated at three
different redshifts.  Here, we simply compute the difference between the
final black hole mass owing to smooth accretion and mergers and the final
black hole mass resulting from smooth accretion alone.  For most black holes,
the contribution from mergers represents less than 10\,\% of the total mass,
with only a small fraction of black holes having merger contributions $>
20$\,\%.  This occurs despite the fact that the inferred mass fraction from
mergers includes some contribution from smooth accretion given by the
relatively increased gravitational torque rates for higher mass black holes
(Equation~(\ref{eq:torque})).  
Interestingly, the mass fraction from mergers
seems to be higher when evaluated for black holes at $z = 2$ relative to either
$z = 4$ or $z = 0$, corresponding to the epoch near the peak of cosmic star formation activity.

Overall, we find that smooth accretion dominates global black hole growth
over cosmic time while black hole mergers may represent a non-negligible
contribution for the most massive black holes at late times, in agreement with
previous studies \citep[e.g.,][]{colberg08,dubois14,kulier13,volonteri13}.
This prediction seems robust for the mass range and redshift range that we
consider here but may be subject to uncertainties relative to the masses
and number densities of seed black holes, the efficiency of black hole
merging during galaxy mergers, or the effects of gravitational recoils
\citep[e.g.,][]{blecha08,bellovary10,bellovary11,micic11}.

\begin{figure} 
\begin{center}
\includegraphics[scale=0.47]{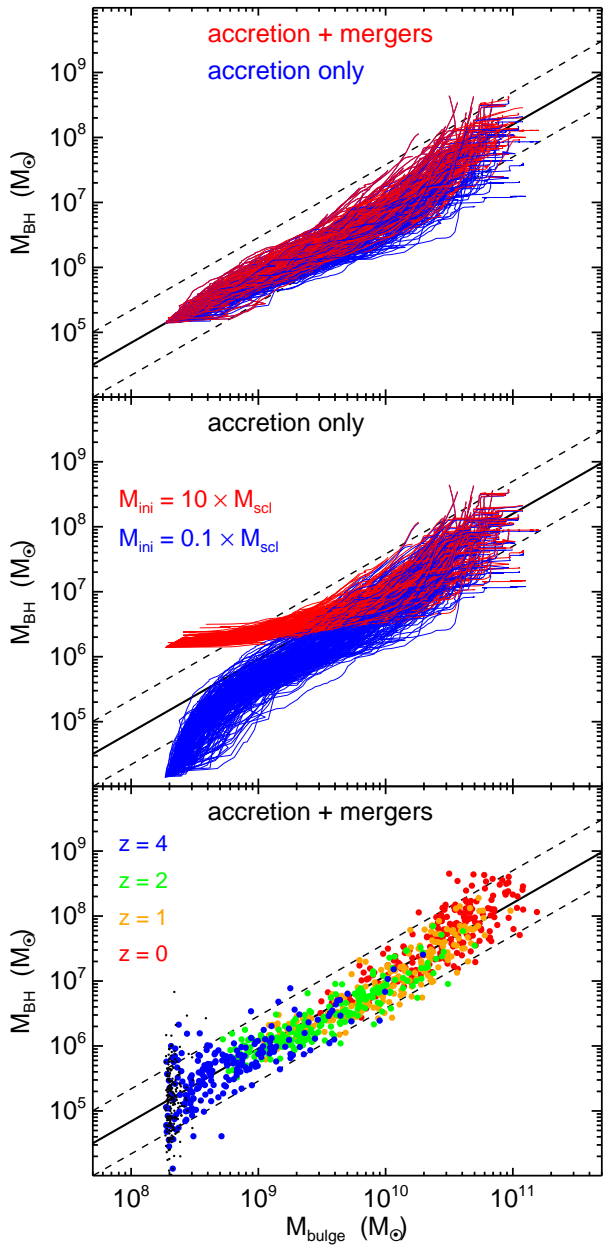} 
\end{center} 
\caption{{\it Top}: evolutionary tracks of galaxies and central black holes in the \Mbulge~plane
for torque-limited growth (blue) and for torque-limited accretion along
with mass contributions from black hole mergers (red).  Black hole seeds
and merging black holes are assumed to lie on the \Mbulge~relation for the
corresponding host galaxy.  The stellar mass within the effective radius is
taken as a proxy for the bulge mass of the host galaxy.  The black solid line
shows the \Mbulge~relation of \citet{har04}; black dashed lines indicate a 0.5
dex scatter in black hole mass.  {\it Middle}: effects of initial conditions
on the black hole--galaxy evolutionary tracks.  We compute torque-limited
growth for seed black holes with initial masses either a factor of 10 above
(red) or below (blue) the corresponding \Mbulge~relation.  In each case, black
holes evolve toward the scaling relation.  {\it Bottom}: \Mbulge~relation at
$z = $ 0 (red), 1 (orange), 2 (green), and 4 (blue) for black holes growing
through torque-limited accretion and mergers.  Masses of black hole seeds
(shown as small black dots) and merging black holes are randomly selected
from a log-normal distribution corresponding to the \Mbulge~relation for
the appropriate galaxy and time step, assuming a 0.5 dex scatter in black
hole mass.} 
\label{fig:Mbulge} 
\end{figure}

\subsection{The $M_{\rm BH}$--$M_{\rm bulge}$ Relation}\label{sec:msig}

In \citet{ang13} we showed that black hole growth by gravitational
torque-driven accretion yields black holes and host galaxies that evolve on
average along the scaling relations from early times down to $z = 2$, provided
that only a small fraction $\epsilon_{\rm m}$ of the inflowing gas feeding onto
the accretion disk from larger scales is finally accreted by the central
black hole.  The mass retention rate $\epsilon_{\rm m} \approx 0.05$ was
found to provide the correct normalization over the full redshift range $z =
8 \rightarrow 2$, assuming that these black holes follow the local
\Mbulge~relation
\citep{har04} for the stellar mass within the effective radius.

Figure~\ref{fig:Mbulge} shows that this result can be extended from the
eight zoom disk galaxies in \citet{ang13} to more than $200$ galaxies in
our cosmological simulation, evolved over a much more extended period of
time from $z \sim 4+\rightarrow 0$.  Provided that the initial conditions
are chosen to agree with the local \Mbulge~relation, black holes and
galaxies grow by more than three orders of magnitude in mass approximately
along the scaling relation, with no further tuning of the mass retention
rate $\epsilon_{\rm m}$.  This conclusion is not affected by the addition of
mass growth from black hole mergers, which was neglected in \citet{ang13}.
The top panel of Figure~\ref{fig:Mbulge} demonstrates this by comparing
the evolutionary tracks in the \Mbulge~plane for black holes growing with
and without contributions from black hole mergers, shown as the red and blue
lines, respectively: the addition of black hole mergers does increase the black
hole masses slightly but does not alter the overall trend.  Smooth accretion
represents most of black hole growth for the majority of the host galaxies
and dominates the overall evolution in the \Mbulge~plane, with black hole
mergers representing typically a small fraction of the total growth.

Since we do not have an a priori reason to assume that seed black
holes correlate with their host galaxy at the starting redshift, we now
examine the impact of relaxing this assumption.  The middle panel of
Figure~\ref{fig:Mbulge} shows the evolutionary tracks predicted by
torque-limited accretion for black hole seeds that are either a factor
of 10 above (red) or below (blue) the \Mbulge~relation by the time their
host galaxies are first resolved in the simulation.  Interestingly, black
holes tend to evolve toward the \Mbulge~relation regardless of the initial
conditions and with no need for mass averaging through mergers or additional
self-regulation processes.  This attractor behavior to lie on
the \Mbulge~relation was
described in \citet{ang13} for a small galaxy sample and it is now confirmed
for a large number of simulated galaxies.  The weak dependence of gravitational
torque rates on black hole mass, namely $\dot{M}_{\rm Torque} \propto M_{\rm
BH}^{1/6}$ (Equation~(\ref{eq:torque})), plays a key role in this overall
convergence process, resulting in a rate at which black holes ``move" in the
logarithmic \Mbulge~plane given by \begin{equation}\label{eq:log} \frac{d}{dt}
\, log(M_{\rm BH}) \; \propto \; \frac{\dot{M}_{\rm BH}}{M_{\rm BH}} \; \propto
\; M_{\rm BH}^{-5/6}, \end{equation} which in turn implies that for a given
host galaxy, a lower (higher) mass black hole grows proportionally faster
(slower) relative to a black hole lying on the \Mbulge~relation.  We will
explore this attractor behavior in more detail in Section~\ref{sec:conv}.

\begin{figure*} 
\begin{center}
\includegraphics[scale=0.45]{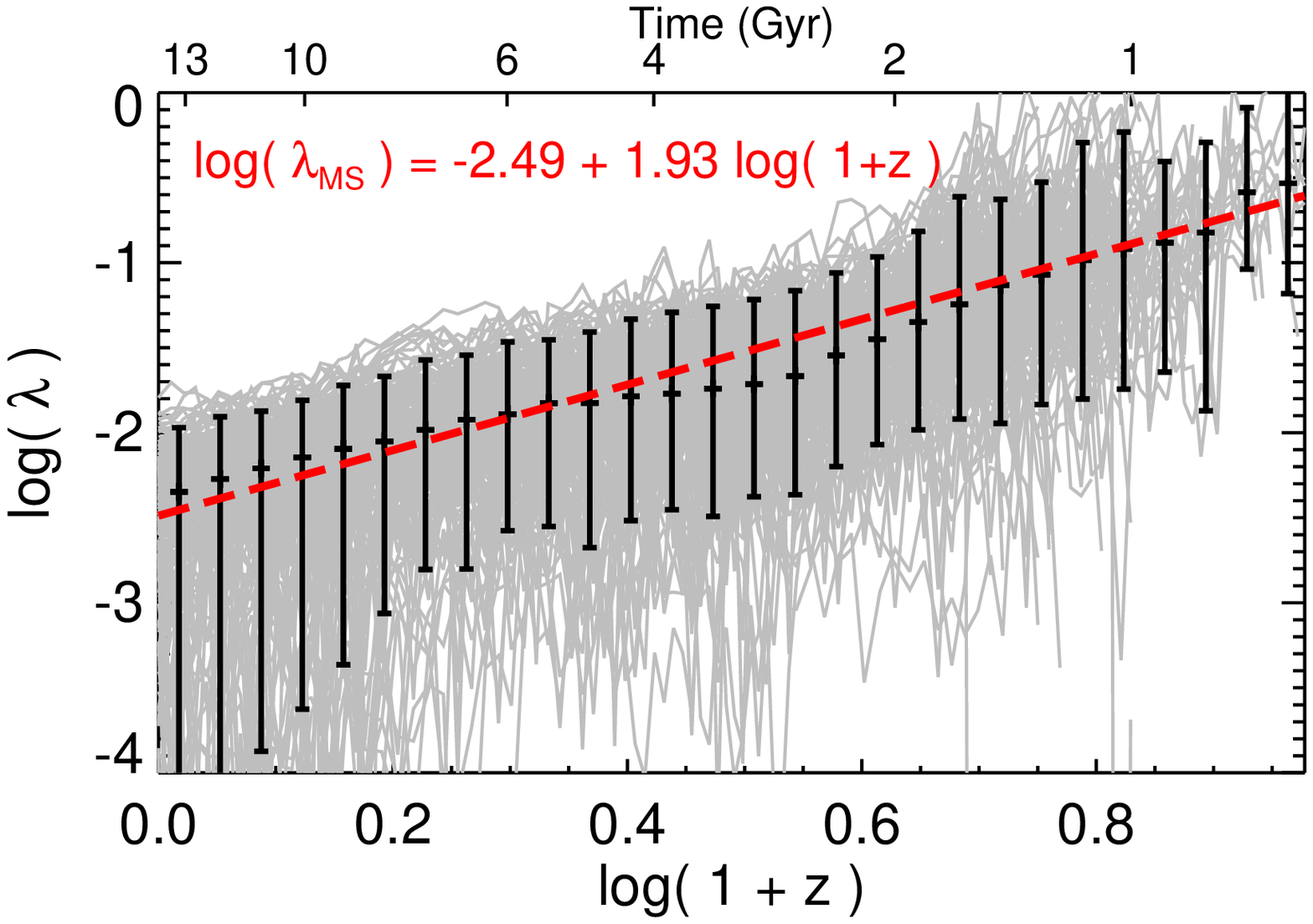}
\includegraphics[scale=0.45]{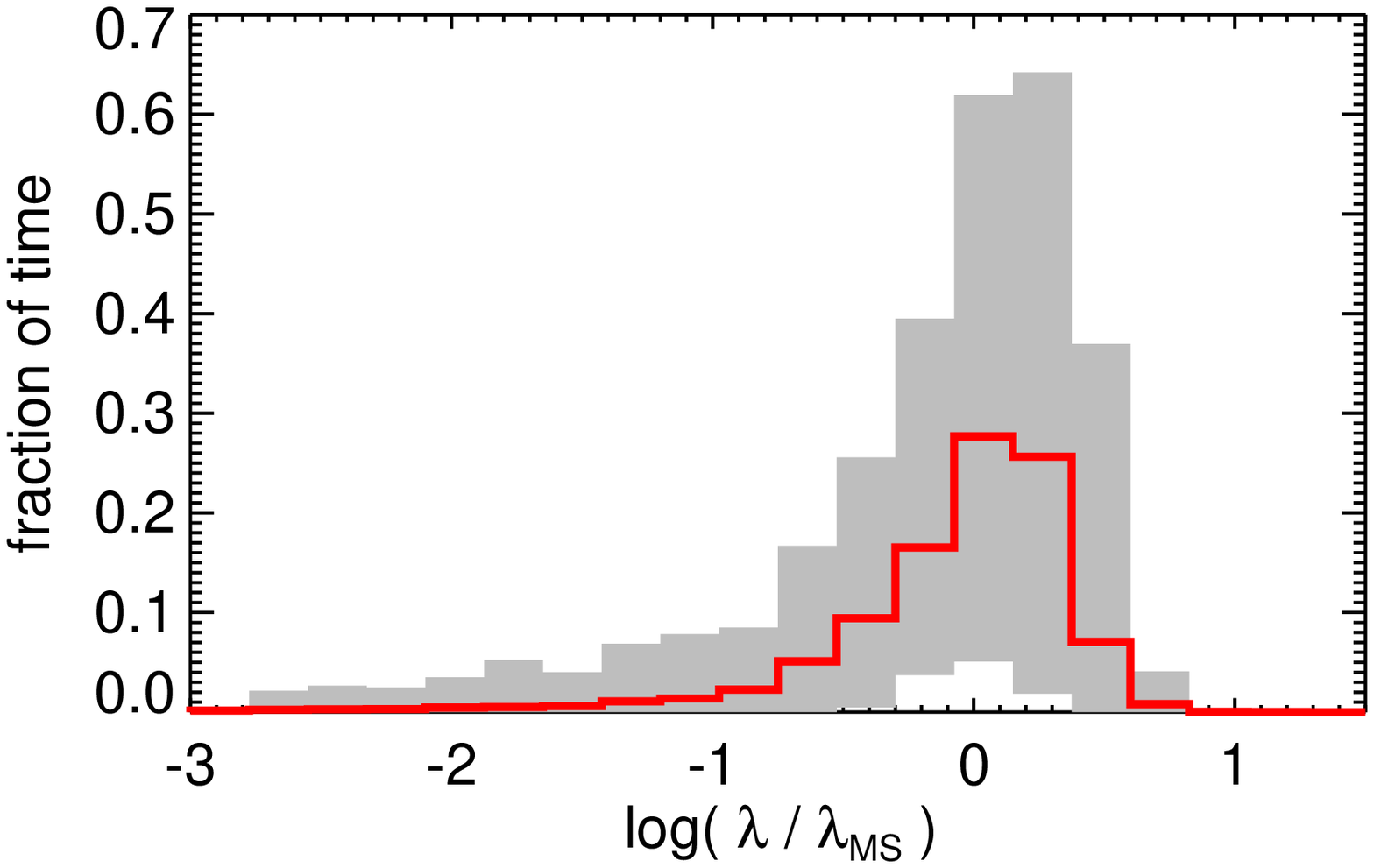} 
\end{center} \caption{{\it Left}: Eddington ratios, $\lambda \equiv \dot{M}_{\rm BH}/\dot{M}_{\rm
Edd}$, as a function of redshift for the central black holes of each of the
\Ngal~galaxies selected at $z = 0$.  Gray lines show individual accretion
histories while black points with error bars show median values within
bins logarithmically spaced in $(1+z)$ and the corresponding 5 and 95
percentiles of the distribution.  Accretion rates are calculated according
to Equations~(\ref{eq:corr}) and~(\ref{eq:torque}) for a mass retention rate
$\epsilon_{\rm m} = 0.05$.  The red dashed line shows the best power law fit to
the median values: ${\rm log}(\lambda_{\rm MS}) \approx -2.49 + 1.93\,{\rm
log}(1+z)$.  {\it Right}: fraction of the evolution time down to $z = 0$
that black holes spend accreting at a given Eddington ratio relative to
$\lambda_{\rm MS}(z)$.  The red solid line shows the time spent in a given
$\lambda / \lambda_{\rm MS}$ bin averaged over all black holes and the gray
shaded region indicates the 5 and 95 percentiles of the distribution of time
fractions in each $\lambda / \lambda_{\rm MS}$ bin.} 
\label{fig:EddRatio}
\end{figure*}

Given that black holes tend to evolve onto the \Mbulge~relation, it seems
justified to adopt the simplification that black holes and galaxies are
already on the scaling relation by the time we define the initial conditions,
i.e., when the host galaxy is first resolved in the cosmological simulation.
Nonetheless, there is significant scatter in black hole mass at a given
bulge mass despite the overall convergence toward the \Mbulge~relation.
This intrinsic scatter does not go away with subsequent evolution; therefore,
it should be taken into account when defining the initial conditions for
black hole growth as well as the mass contribution from black hole mergers.

The bottom panel of Figure~\ref{fig:Mbulge} shows the \Mbulge~relation
obtained at different redshifts when black hole seeds and merging black
holes are randomly chosen from a log-normal distribution corresponding to
the \Mbulge~relation for the appropriate galaxy and time step, but
assuming a 0.5
dex scatter in black hole mass.  Overall, our full model for torque-limited
growth is consistent with a close-to-linear non-evolving \Mbulge~relation,
so long as the initial conditions at some reference redshift are not biased
toward either higher-mass or lower-mass black holes relative to their host
galaxies.  Note that some initially large log-normal scatter may produce a
bias toward higher-mass black holes at later times because their time-scale
for convergence toward the scaling relation is significantly longer relative
to lower-mass black holes.  We explore this in Section~\ref{sec:conv}.

Black hole mergers may reduce the scatter of the \Mbulge~relation by
recurrent mass averaging \citep{hirschmann10}, a process that has indeed
been suggested as the actual physical mechanism giving rise to the black
hole--galaxy scaling relations \citep{pen07,jahnke11}.  Mergers actually
seem to reduce the scatter somewhat (Figure~\ref{fig:Mbulge}, top panel),
but major mergers ($>$1:5) are clearly not frequent enough for our galaxy
sample to establish the \Mbulge~relation in the first place.  In some
cases, the merging of several slightly over-massive black holes may yield
outliers in the \Mbulge~relation even under normal accreting conditions.
It is, nonetheless, challenging to explain recent observations in the local
universe suggesting the presence of highly over-massive black holes compared
to their host galaxies (\citealp{bog12,vdb12}; but see \citealp{emsellem13}).
In the context of torque-limited growth, it is plausible that such extreme
objects could form from highly above-average accreting conditions, such as
a favorably oriented galaxy merger.

Observations are currently inconclusive regarding the slope and
normalization of the scaling relations at high redshift.  While
several studies have reported an increase in the black hole mass to
host galaxy mass ratio for individual systems at higher redshifts
\citep{treu07,decarli10,greene10,merloni10,bennert11,targett12}
there remain significant concerns about to systematics in the mass
estimators \citep[e.g.,][]{park13} and biases introduced by selection effects
\citep{lauer07,shen10,schulze11}.  Indeed, a number of observations seem
consistent with little or no evolution in the black hole mass to host galaxy
mass ratio \citep{Jahnke09,cisternas11a,schramm13}.  Assuming a non-evolving
mass retention rate ($\epsilon_{\rm m}$) in the accretion flow, torque-limited
growth predicts no significant evolution of the \Mbulge~relation unless the
initial conditions are substantially different relative the local scaling
relation.  In Section~\ref{sec:conv}, we evaluate the characteristic time
scales for convergence toward the \Mbulge~relation.

Note that we have not attempted to estimate the ``true" bulge
mass in analogy with observations, but instead replaced it in the
\Mbulge~relation by the stellar mass within the effective radius of
the host galaxy.  Torque-limited growth yields a correlation between
black hole mass and stellar mass regardless of the morphology of the
galaxy.  This suggests that the processes driving the morphological
evolution of the stellar component in galaxies may not be fundamental
for the growth of their central black hole \citep{marleau13,simmons13}.
Incidentally, there is increasing evidence for significant black hole
growth taking place in disk dominated galaxies with no merger signatures
\citep{gabor09,georgakakis09,cisternas11b,kocevski12,mull12b,schawinski12,treister12}, 
while both, galaxy mergers and secular evolution, are commonly invoked as primary mechanisms for bulge formation \citep[e.g.,][]{hopbun10,kormendy13}.

This simple scenario of black hole--galaxy coevolution is challenged by
observations in the local universe suggesting that black holes correlate
differently with different galaxy components \citep{gra08,hu08,gra11,kormendy11,kormendy13}.
Recent results imply a broken power-law relation between the masses of black
holes and their host spheroids \citep{gra12,gra13,scott13}, with lower-mass
black holes in S\'ersic galaxies ($M_{\rm BH} \lesssim 10^8$\,\Msun) following
a steeper relation $M_{\rm BH} \propto M_{\rm bulge}^2$ below the classic
nearly linear scaling.  While our simulations lack the resolution required
for a detailed analysis of the $z = 0$ \Mbulge~relation and its morphological
dependence, we note that torque-limited growth yields a qualitatively similar
steep trend for initially under-massive black holes as they evolve onto the
\Mbulge~relation (Figure~\ref{fig:Mbulge}, middle panel).  Observations of
black holes in low-mass galaxies may thus provide significant constraints
on the initial conditions for massive black hole growth \citep{greene12}.

\subsection{Evolution of Eddington Ratios}\label{sec:Edd}

Gravitational torques drive gas inflows from galactic scales down to sub-parsec
scales, feeding the accretion flow near the black hole, and governing the
co-evolution of black holes and galaxies.  The observed black hole-galaxy
scaling relations are a natural outcome of this process.  In this section,
we explore the accretion histories resulting from torque-limited growth as
well as implications for observations of active systems across cosmic time.

The left panel of Figure~\ref{fig:EddRatio} shows the evolution of Eddington
ratios with redshift, defined here as the black hole accretion rate in units
of Eddington, $\lambda \equiv \dot{M}_{\rm BH} / \dot{M}_{\rm edd}$.  For all
black holes, $\dot{M}_{\rm BH}$ is calculated from Equation~(\ref{eq:corr})
for the mass retention rate $\epsilon_{\rm m} = 0.05$.  The Eddington rate
is given by the usual definition, $ \dot{M}_{\rm edd} = 4 \pi G M_{\rm
BH} m_{\rm p} / (\eta \sigma_{\rm T} c)$, where the accretion efficiency,
$\eta$, represents the maximum amount of potential energy per unit rest mass
energy that can be extracted from the innermost stable circular orbit of
the accretion disk around the black hole.  Throughout this paper, we adopt
a fixed value $\eta = 0.1$ \citep[e.g.,][]{yu02,marconi04} and ignore its
intrinsic dependence on black hole spin.

Gray lines in Figure~\ref{fig:EddRatio} (left panel) correspond to the
accretion histories of individual black holes.  Despite our limited time
resolution, restricted by the number of output files produced during
the simulation, accretion rates show significant variability relative to
cosmological timescales.  This variability follows from the complex evolution
of the inner regions of galaxies \citep{hop10}, which manifests itself in
the gravitational torque model as significant variations in morphological
properties within the radial aperture $R_{0}$ \citep{hop11}.  Black points
with error bars show median Eddington ratios within logarithmically spaced
bins in $1+z$ and the 5 and 95 percentiles of the distribution, indicating
that there is also a significant scatter for our sample of black holes at any
given redshift.

Despite the large scatter, our simulations reveal a common trend for
the evolution of Eddington ratios.  Black holes are typically accreting
at high Eddington ratios at early times, with median values $\lambda >
10$\,\% at $z \gtrsim 6$ and may even exceed the Eddington limit in
some cases.  At lower redshifts, a gradual decrease in Eddington ratios
yields $\lambda \sim 1$--10\,\% at $z \approx 2$ \citep[as previously
found in][]{ang13}, reaching typical present day values $\lambda \sim
0.1$--1\,\% at $z = 0$.  As shown by the red dashed line in the left panel
of Figure~\ref{fig:EddRatio}, a simple power law provides a good fit to the
redshift dependence of the median Eddington ratio, albeit with significant
scatter: \begin{equation}\label{eq:Edd} {\rm log}( \lambda_{\rm MS} ) \approx
-2.49 + 1.93\,{\rm log}( 1+z ), \end{equation} where we have ignored any
intrinsic dependence of Eddington ratios on black hole mass (see below).
The exact slope and normalization in Equation~(\ref{eq:Edd}) are somewhat
dependent on sample selection and initial conditions (Section~\ref{sec:conv}).
Nonetheless, this relation provides a useful tool for characterizing black
hole accretion histories, in analogy with the star formation main sequence,
which can be defined in terms of the median specific SFR for a given redshift
interval \citep[e.g.,][]{dav11b,elbaz11}.

We can now evaluate the evolution of Eddington ratios relative to the sequence
defined by Equation~(\ref{eq:Edd}).  The right panel of Figure~\ref{fig:EddRatio}
shows the fraction of time that black holes spend accreting at a given
Eddington ratio in units of the median value $\lambda_{\rm MS}$.  For each
black hole at a given redshift, we calculate the ratio $\lambda(z) /
\lambda_{\rm MS}(z)$ to which we assign the duration of the current time
step.  Then, by adding up the contributions from all time steps, we estimate
the fraction of the total evolution time (down to $z = 0$) during which a
given black hole grows at some Eddington ratio relative to the main
sequence value.  We indicate as the red solid line the average fraction of
time spent in a given $\lambda / \lambda_{\rm MS}$ bin over all black holes
(equivalent to the probability per logarithmic interval), while the gray
shaded region corresponds to the 5 and 95 percentiles of the distribution
in each $\lambda / \lambda_{\rm MS}$ bin.

The right panel of Figure~\ref{fig:EddRatio} shows that black holes spend
most of their time accreting near the median Eddington ratio for the whole
population, suggesting that Equation~(\ref{eq:Edd}) may, indeed, represent
an ``AGN main sequence" \citep{mull12a}.  Eddington ratios can be roughly
described by a log-normal distribution centered at $ \lambda_{\rm MS}(z)$ at
all redshifts, but note the asymmetry with respect to $\lambda = \lambda_{\rm
MS}$, with a relative increased probability for black holes accreting at
lower Eddington ratios (especially at low redshift).  One caveat here is the
limited time resolution; our inferred Eddington ratios correspond to
average values within time intervals ranging from $\sim 10$ to 300\,Myr in the
redshift range $z \sim 6 \rightarrow 0$, while AGN luminosities exhibit strong
variability over a large dynamic range, from hours \citep[e.g.,][]{mchardy13}
to Myr time scales \citep[e.g.,][]{mcnamara07,goncalves08}.  Thus, the right
panel of Figure~\ref{fig:EddRatio} corresponds to departures from the AGN main
sequence ($\lambda_{\rm MS}$) on timescales comparable to typical galaxy
dynamical timescales.  Shorter timescale variability that we cannot track
may have important consequences for the observed distribution of Eddington
ratios and the inferred connection between star formation and AGN activity
\citep{hickox14}.

\begin{figure} 
\begin{center}
\includegraphics[scale=0.45]{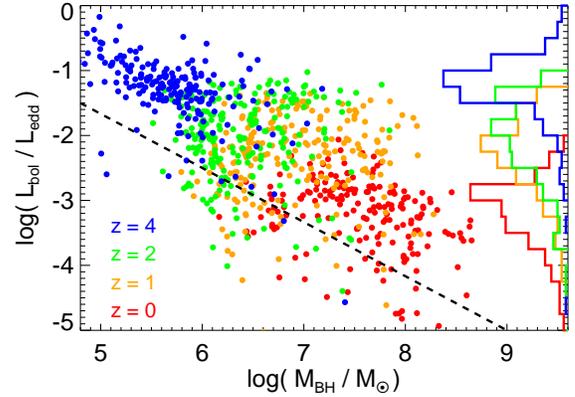} 
\end{center} 
\caption{Bolometric luminosity in units of the Eddington luminosity, $L_{\rm bol}/L_{\rm edd}$,
as a function of black hole mass for our sample of \Ngal~black holes at $z = $ 0 (red), 1
(orange), 2 (green), and 4 (blue).  Bolometric luminosities are calculated from
accretion rates by assuming that there is a transition between radiatively
efficient to radiatively inefficient accretion at $\lambda_{\rm crit} =
0.03$ \citep{merloni08}.  Histograms show the distribution of Eddington ratios
at each redshift (arbitrarily normalized).  The black dashed line corresponds
to the scaling $\lambda \propto M_{\rm BH}^{-5/6}$.} 
\label{fig:LratiovsMbh}
\end{figure}

\subsection{Bolometric Luminosities}\label{sec:Lbol}

The radiative properties of accretion flows around AGNs are thought to
depend primarily on the mass inflow rate onto the black hole, with a
relatively well defined transition between radiatively efficient and
radiatively inefficient modes at Eddington ratios of about a few percent
\citep{narayan95b,maccarone03,greene06}, in close analogy to Galactic
stellar-mass black holes in X-ray binaries \citep{remillard06}.  Here, we
infer AGN bolometric luminosities by assuming that there is an accretion
state transition at $\lambda_{\rm crit} = 0.03$, as in \citet{merloni08}.
For the radiatively efficient mode ($\lambda > \lambda_{\rm crit}$), the
bolometric luminosity is simply proportional to the accretion rate, $L_{\rm
bol} = \eta \,\dot{M}_{\rm BH}\,c^2$, and, therefore,  $L_{\rm bol}/L_{\rm
edd} = \lambda$.  For radiatively inefficient accretion flows ($\lambda
< \lambda_{\rm crit}$) we compute $L_{\rm bol} = \eta \, \dot{M}_{\rm
BH}\,c^2\,(\lambda/\lambda_{\rm crit})$.

Figure~\ref{fig:LratiovsMbh} shows the ratio of the bolometric luminosity
to the Eddington luminosity, $L_{\rm bol}/L_{\rm edd}$, as a function of
black hole mass, evaluated at four different redshifts.  With the definition
of $\lambda_{\rm crit}$ adopted here, the shift from radiatively efficient
accretion to radiatively inefficient accretion occurs at $z \approx 1$--2 for
the median black hole.  Thus, the probability of a galaxy hosting an AGN in
the radiatively inefficient mode increases at lower redshifts.  The inferred
bolometric luminosity corresponding to a given accretion rate is lower for
black holes growing in the radiatively inefficient mode.  The ratio $L_{\rm
bol}/L_{\rm edd}$ is, therefore, characterized by a stronger evolution with
redshift relative to the intrinsic accretion rate ($\lambda$), decreasing
by about three orders of magnitude from $z = 4$ to $z = 0$.

At a fixed redshift, the inferred $L_{\rm bol}/L_{\rm edd}$ values roughly
follow a log-normal distribution, as expected from Figure~\ref{fig:EddRatio}.
At $z = 4$, most black holes radiate in a relatively narrow log-normal
distribution around $L_{\rm bol}/L_{\rm edd} \approx 0.05$, right above
$\lambda_{\rm crit}$.  At lower redshifts, the width of the $L_{\rm bol}/L_{\rm
edd}$ distribution increases, with an extended tail toward low luminosities.
This is particularly evident at $z \approx 1$--2 when most black holes are
undergoing a transition from radiatively efficient ($\lambda > \lambda_{\rm
crit}$) to radiatively inefficient ($\lambda < \lambda_{\rm crit}$) accretion.
Note that the absence of black holes with masses $>10^8$\,\Msun~at $z\sim 2$ simply reflects the lack of massive galaxies early enough to host such massive black holes in our simulated volume.

Figure~\ref{fig:LratiovsMbh} shows an anti-correlation between $L_{\rm
bol}/L_{\rm edd}$ and black hole mass.  For a given host galaxy, the
gravitational torque rate scaled by the Eddington rate
is lower for higher-mass black
holes, $\lambda \propto M_{\rm BH}^{-5/6}$.  This should imply a strong
trend of decreasing Eddington ratios for increasing black hole mass at a fixed
redshift; this trend is, however, weaker than expected for our sample of black
holes or even nonexistent at $z \approx1$--2, suggesting a complex evolution
of black hole accretion besides the intrinsic dependence on $M_{\rm BH}$.
Indeed, we will show in Section~\ref{sec:resT} (Figure~\ref{fig:res}) that
higher mass galaxies are more compact than lower mass galaxies while having a
similar disk fraction; this results in higher values of $R_{0}^{-3/2}M_{\rm d}$
that partially compensate for the decrease in $\lambda$ values with increasing
$M_{\rm BH}$.  The net effect is a $\lambda(M_{\rm BH})$ dependence that is
weaker
than the a priori expected $\lambda \propto M_{\rm BH}^{-5/6}$, which may
only be identified for a black hole sample spanning a sufficient mass range.

\begin{figure*} 
\begin{center}
\includegraphics[scale=0.4]{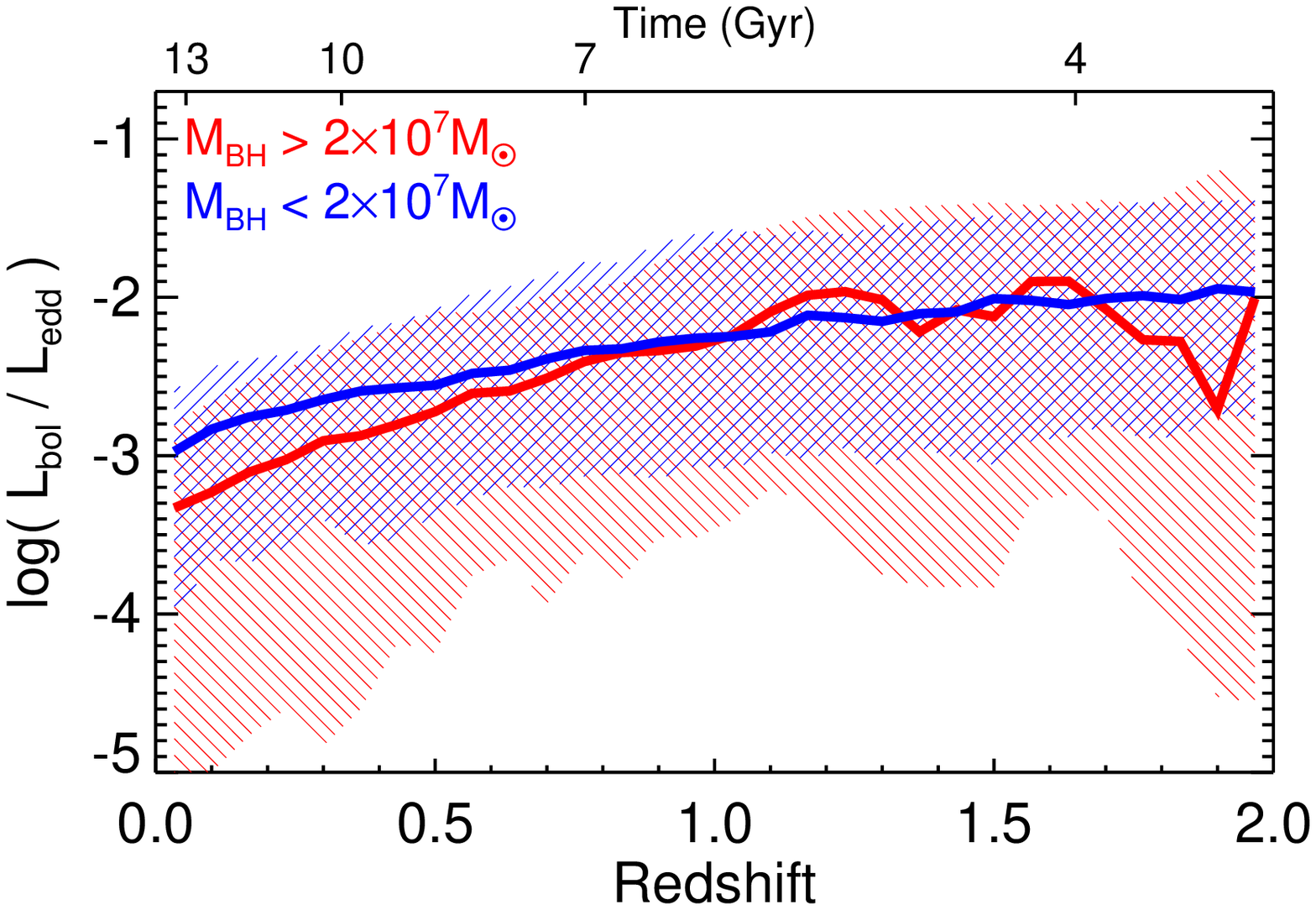}
\includegraphics[scale=0.4]{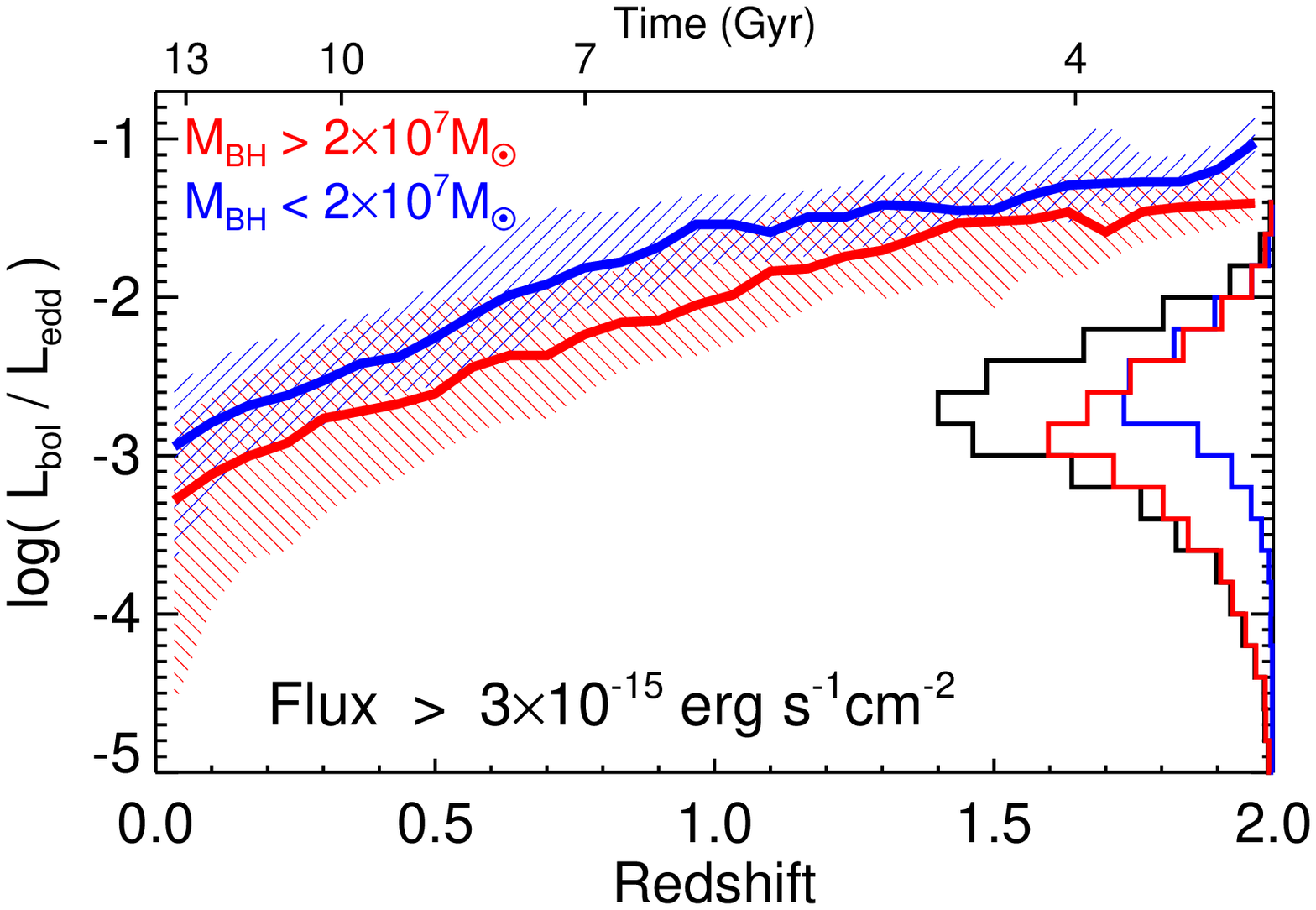} 
\end{center} 
\caption{Impact of selection effects on the inferred evolution and mass dependence of
Eddington ratios.  {\it Left}: bolometric luminosity in Eddington units as
a function of redshift from $z = 2$ down to $z = 0$ for the sub-samples of
black holes with masses above (red) and below (blue) $2 \times 10^7$\,\Msun~at
each redshift.  Red and blue solid lines show median values for the high-mass
and low-mass samples, respectively, while the red and blue hatched areas
correspond to the central 75\,\% of the distribution for the population
of black holes in each sub-sample.  {\it Right}: same as the left panel
but only including black holes radiating above a total flux limit $F_{\rm
lim} = 3 \times 10^{-15}$\,erg\,s$^{-1}$\,cm$^{-2}$, which corresponds to a
bolometric luminosity $L_{\rm bol} \approx 10^{44}$\,erg\,s$^{-1}$ at $z = 2$.
Histograms show the distribution of Eddington ratios (arbitrarily normalized)
for all black holes below $z < 0.5$ (black) and for the high-mass (red)
and low-mass (blue) sub-samples.} 
\label{fig:Lratio_z4} 
\end{figure*}

Selection effects may have a significant impact on the observed evolution
of Eddington ratios and their dependence on black hole mass.  This is
illustrated in Figure~\ref{fig:Lratio_z4} where we calculate the median
bolometric luminosity in Eddington units ($L_{\rm bol}/L_{\rm edd}$) as
a function of redshift ($z = 2 \rightarrow 0$) for two sub-samples of
black holes: those with masses above and below
$M_{\rm BH} = 2 \times 10^7$\,\Msun.  The left panel of 
Figure~\ref{fig:Lratio_z4}, which includes all of our \Ngal~black
holes selected at $z = 0$, shows no clear separation of the high-mass and
low-mass sub-samples in terms of their median $L_{\rm bol}/L_{\rm edd}$.
The high-mass sample is characterized by larger scatter at all redshifts,
perhaps suggesting a stronger variability, but the limited mass range does
not allow us to discern an intrinsic decrease in $\lambda$ with $M_{\rm BH}$.

The right panel of Figure~\ref{fig:Lratio_z4} mimics the effects of 
observational sensitivity limits by calculating median Eddington ratios for
the two sub-samples but now only including black holes radiating with
a total ``observed" flux higher
than $F_{\rm lim} = 3 \times 10^{-15}$\,erg\,s$^{-1}$\,cm$^{-2}$, equivalent
to a bolometric luminosity $L_{\rm bol} \approx 10^{44}$\,erg\,s$^{-1}$ at $z
= 2$.  For each black hole we calculate the total flux from the bolometric
luminosity as $F = L_{\rm bol} / (4 \pi d_{\rm L}^2)$, where $d_{\rm L}$
is the luminosity distance at the corresponding redshift.  Our flux-limited
sample of black holes shows a clear dependence of Eddington ratios on black
hole mass for the full redshift range $z = 2 \rightarrow 0$, with the higher
mass sample dominating the low $L_{\rm bol}/L_{\rm edd}$ regime.  This simple
experiment illustrates how the inferred evolution of black hole populations
can be affected by sensitivity limits, even neglecting obscuration effects.
Black holes growing at low Eddington ratios may be missed in flux-limited
surveys preferentially at higher redshifts and for lower-mass black holes, 
as explicitly shown by \citet{kollmeier06}.
This may result in (1) a stronger apparent evolution of Eddington ratios
with redshift and (2) an artificially increased systematic offset
between the typical Eddington ratios of higher-mass and lower-mass black holes.

A direct comparison of the evolution of Eddington ratios with observations
is not trivial given that differences in, for example, selection techniques and
completeness limits often yield contrasting results among AGN populations.
Despite this, a wide range of observations indicate that the typical
Eddington ratios of accreting black holes increase at higher redshifts
\citep{kollmeier06,netzer07,kauffmann09,derosa11,trak11,aird12,bongiorno12,lusso12,trak12},
in broad agreement with our results and consistent with expectations from
AGN synthesis models \citep[e.g.,][]{merloni08,shankar13}.  However, the
typical median values may vary significantly for different AGN populations
even at similar redshifts, e.g., ranging from near Eddington accretion
for the AGN sample of \citet{kollmeier06} at $z \sim 0.3$--4 to sub-Eddington
growth for the AGN population studied by \citet{lusso12} at $z \sim 1$--2.
Indeed, \citet{trump11} reported the presence of two separate populations in
X-ray-selected AGNs, associated with black holes accreting in radiatively
efficient ($\lambda \gtrsim 0.01$) and radiatively inefficient ($\lambda
\lesssim 0.01$) modes.  For torque-limited growth, such a transition between
accretion modes occurs multiple times for typical black hole accretion
histories, but with an increased probability for radiatively inefficient
accretion at lower redshifts.

The situation is less clear with respect to the distribution of
Eddington ratios and its dependence on black hole mass.  Numerous
observational studies have reported Eddington ratio distributions
consistent with a log-normal distribution for a wide range of redshifts
\citep{kollmeier06,netzer07,hickox09,netzer09,trak11,trump11,lusso12}
in agreement with our results, while other studies favor a universal power
law distribution function independent of black hole mass \citep{aird12,bongiorno12}.  
\citet{trak12} reported a significant decrease
in Eddington ratios with increasing black hole mass in the redshift range
$z = 0$--2, while \citet{kelly13} found that $\lambda$ is approximately
independent of black hole mass at low ($z < 0.8$) and high ($z > 2.65$)
redshifts but increases with black hole mass at intermediate redshifts.
\citet{kollmeier06} examined the distribution of Eddington ratios in the redshift range $z \sim 0.3$--4 and found a characteristic log-normal distribution independent of black hole mass and redshift down to well-characterized completeness limits. 
At low redshift, \citet{kauffmann09} identified two distinct distributions of Eddington
ratios: black holes in star forming galaxies follow a
log-normal distribution that only weakly depends on the black hole mass
and black holes in passive galaxies follow a power-law distribution function with a normalization
that strongly depends on the black hole mass.

We find that Eddington ratios averaged over galaxy evolution timescales can
be roughly described by a lognormal distribution with increasing width at
lower redshifts and with a median value evolving as a power law in $(1+z)$
broadly similar to the cosmic evolution of specific SFRs.  Furthermore,
the combined dependence of accretion rates on black hole mass and galaxy
surface density ($R_{0}^{-3/2}M_{\rm d}$) yields a weak trend of decreasing
median Eddington ratios with increasing black hole mass at all redshifts.
Encouragingly, similar trends have been identified by \citet{shankar13} as
key elements in reproducing a number of observations, including the observed
Eddington ratio distributions, the high AGN fractions at low redshift,
and the higher frequency of AGNs in higher-mass galaxies.

\subsection{The SFR--AGN Connection}\label{sec:sfagn}

Torque-limited accretion yields black holes growing, on average, in tandem with
their host galaxies (Figure~\ref{fig:Mbulge}).  Smooth accretion dominates
the total growth of black holes (Figure~\ref{fig:hist_mrg}) and their host
galaxies \citep[e.g.,][]{mura02,ker05}, implying that there must be some
connection between the total SFR of galaxies and their nuclear activity on
cosmological timescales.  In this section, we extend our current analysis
to a significantly larger number of black holes and host galaxies 
to present predictions for the relation between galaxy SFRs and AGN activity
for an increased dynamic range.

For this purpose, we select all galaxies with stellar masses $M_{\rm star}
> 10^8$\,\Msun~from our $[32\,\hmpc]^3$ simulation volume at different
redshifts, for example there is a total of 4356 and 5815 galaxies at $z=0$ and
$z = 2$, respectively.  At each redshift, we assign central black holes to every
galaxy by assuming consistency with the local \Mbulge~relation \citep{har04};
black holes are randomly selected from a log-normal distribution centered
on the \Mbulge~relation for each galaxy and assuming a 0.5 dex scatter
in black hole mass.  We estimate accretion rates by direct evaluation of
Equation~(\ref{eq:corr}) with $\epsilon_{\rm m} = 0.05$, where we now employ
a radial aperture equal to the effective radius of the host galaxy, $R_{0}
= R_{\rm eff}$ (see Equation~(\ref{eq:torque})).  This allows us to evaluate
accretion rates for virtually all galaxies within our simulation volume.
The effects of using different radial apertures in the gravitational torque
model are discussed in Section~\ref{sec:res}.

\begin{figure} 
\begin{center}
\includegraphics[scale=0.5]{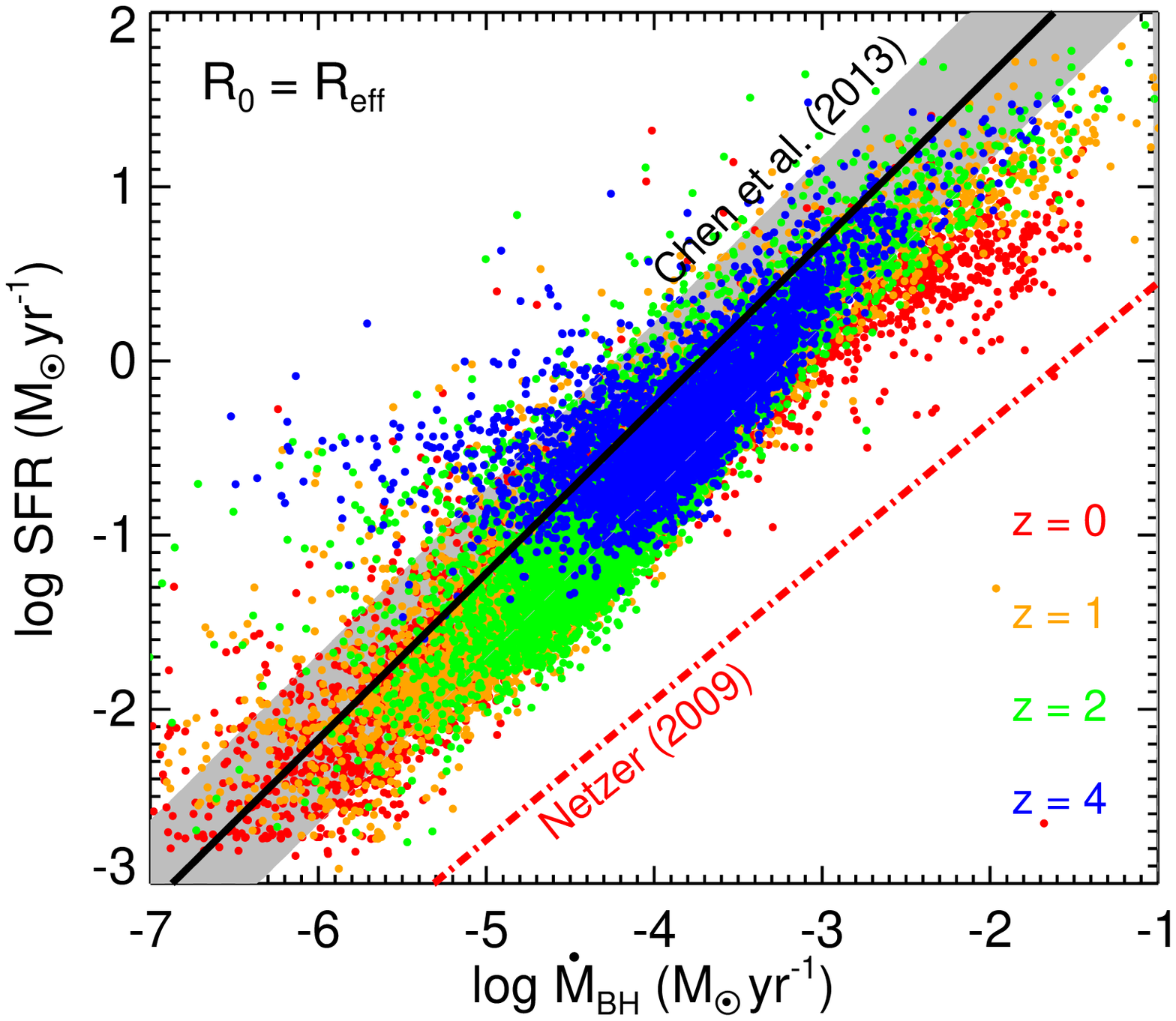} 
\end{center} 
\caption{Total
SFR as a function of central black hole accretion rate for all galaxies with
stellar masses $M_{\rm star} > 10^8$\,\Msun~found within our $[32\,\hmpc]^3$
simulation volume at $z = $ 0 (red), 1 (orange), 2 (green), and 4 (blue).
Accretion rates are calculated by direct evaluation of Equation~(\ref{eq:corr})
(with $\epsilon_{\rm m} = 0.05$), assuming that black holes lie on the
\Mbulge~relation \citep[][for the stellar mass within the effective
radius]{har04} with a 0.5 dex scatter in black hole mass at all redshifts.
Torque-limited inflow rates (Equation~(\ref{eq:torque})) are calculated
from host galaxy properties evaluated within a radial aperture equal to
the effective radius, $R_{0} = R_{\rm eff}$.  The black solid line shows
the SFR--$\dot{M}_{\rm BH}$ correlation reported by \citet{chen13} for
star-forming galaxies in the redshift range $0.25 < z < 0.8$, and the grey
shaded region corresponds to their estimated uncertainty in the normalization.
The red dot--dashed line shows the $L_{\rm SF} \propto L_{\rm AGN}^{0.8}$
relation of \citet{netzer09} for low redshift AGN-dominated systems.}
\label{fig:SFRvsMdot} 
\end{figure}

Figure~\ref{fig:SFRvsMdot} shows the SFR--$\dot{M}_{\rm BH}$ relation predicted
by the gravitational torque model.  Points with different colors represent
the location of individual systems in the SFR--$\dot{M}_{\rm BH}$ plane at
different redshifts, from $z = 4$ (blue) to $z = 0$ (red).  Despite the large
scatter, the increased dynamic range allows us to identify a clear relation
extending over a few orders of magnitude in both SFR and $\dot{M}_{\rm BH}$;
the inferred slope resembles the close-to-linear scaling expected from
the local \Mbulge~relation.  Note that the overall normalization of the SFR
relative to $\dot{M}_{\rm BH}$ is controlled by the mass retention rate, which
is set to $\epsilon_{\rm m} = 0.05$ to match the \Mbulge~relation
of \citet{har04} for the stellar mass within the effective radius of the
host galaxy (i.e., for one-half of the total stellar mass).

Torque-limited growth yields, therefore, a connection between AGN activity
and SFR on timescales comparable to the dynamical timescale of galaxies.
Cosmological gas infall and transport of angular momentum in the galaxy by
gravitational instabilities appear to be the primary physical drivers behind
this relation.  The evolution of specific SFRs with redshift can be roughly
described by a characteristic power law given by the evolution of the
gas accretion
rate onto dark matter halos, $\sim (1 + z)^{2.25}$ \citep{dek09}; feedback
can modify this by suppressing star formation at early epochs and providing
recycling wind material back onto galaxies at later times, yielding a shallower
evolution of the specific SFR with redshift \citep{opp10,dav11b,ang14}.
Intriguingly, gravitational torques provide gas inflows for fuelling the
central AGN at a roughly constant fraction of the SFR in galaxies, in a time
averaged sense.  The evolution of Eddington ratios resulting from this process
can also be described as a power law in $1 + z$ (Figure~\ref{fig:EddRatio}),
suggesting that central black holes evolve along an AGN main sequence similar
to the main sequence for star-forming galaxies \citep{mull12a}.
Note that this is in contrast to recent interpretations of the SFR--AGN connection in terms of positive AGN feedback triggering star formation \citep[e.g.,][]{silk13,zubovas13}.

The SFR--$\dot{M}_{\rm BH}$ connection is, however, not direct on a
galaxy-by-galaxy basis at all times.  Even at the scales resolved here, the
accretion rates are highly variable (Figure~\ref{fig:EddRatio}) and may
experience significant variations uncorrelated with the host galaxy SFR
\citep{hop10,hop11,ang13}.  Indeed, the evolutionary tracks of individual
systems in the SFR--$\dot{M}_{\rm BH}$ plane are rather complicated.  In our
simulations, star formation responds to variations in gas surface density
via a sub-grid prescription tuned to match the observed \citet{ken98}
relation \citep{spr03}.  On average, gas inflows by gravitational torques
also increase with gas surface density via the $R_{0}^{-3/2}M_{\rm d}$ term
in Equation~(\ref{eq:torque}) (as well as the intrinsic dependence on $f_{\rm
gas}$), but critically depend on the overall morphology of the inner region of
the galaxy.  Variations in the fraction of mass in a disk component ($f_{\rm
d}$) are responsible for significant scatter in the SFR--$\dot{M}_{\rm BH}$
relation.  Note that we have assumed a fixed mass retention rate $\epsilon =
0.05$; relaxing this assumption could result in additional scatter.

Do we observe such a correlation between SFR and AGN activity?
Figure~\ref{fig:SFRvsMdot} shows substantial agreement between our results
and recent observations by \citet{chen13}.  These authors showed that the
average central black hole accretion rate for star-forming galaxies in
the redshift range $0.25 < z < 0.8$ is almost linearly proportional to the
SFR; their inferred SFR--$\dot{M}_{\rm BH}$ relation is shown as the black
solid line, and the uncertainty in the normalization is indicated by the
gray shaded region.  A roughly similar correlation between star formation
and AGN luminosity was reported by \citet{netzer09} for active galaxies at
lower redshifts,  $L_{\rm SF} \propto L_{\rm AGN}^{0.8}$, which we show as
the red dot--dashed line in Figure~\ref{fig:SFRvsMdot}.  In this case, the
normalization is significantly lower, perhaps unsurprisingly given that this
relation was found for AGN-dominated systems.  Note that we are extending the
correlations of \citet{netzer09} and \citet{chen13} to a SFR--$\dot{M}_{\rm
BH}$ regime well below their detection limits.  We predict that similar
correlations should continue down to significantly lower levels of star
formation and black hole accretion.

Positive correlations between average SFR and AGN luminosity have also been
reported by a number of authors at higher redshifts \citep{feltre13},
in general agreement with our findings, but seem to hold only for the
highest luminosity systems \citep{lutz10,rosario12,rovilos12}.  Other studies
suggest little or no connection between the average SFR and black hole
accretion \citep{harrison12} or even link luminous AGN activity with a 
suppression of star
formation \citep{page12}, in apparent contradiction with torque-limited growth.
However, as recently discussed by \citet{hickox14}, AGN variability may have
important consequences for the observed SFR--$\dot{M}_{\rm BH}$ correlations.

Global changes in star formation occur on timescales comparable to the
dynamical timescale of the galaxy while high resolution simulations show
that significant AGN variability may occur at essentially all timescales
\citep{hop10,levine10,nov11,gabor13}.  Furthermore, SFR tracers are typically
sensitive to star formation events on timescales up to $\sim 100$\,Myr, while
the measurements of, for example,
X-ray luminosity in AGN track the ``instantaneous"
black hole accretion rate.  Therefore, a direct connection between SFR and
AGN activity is only expected when one averages black hole accretion rates over
sufficiently long time periods (which is obviously not possible in the
observations)
or, alternatively, when one calculates the average ``instantaneous" accretion
rate as a function of host galaxy SFR for a statistical sample of systems.
Indeed, this second approach has been pursued recently with results that
strongly suggest a co-evolution of star formation and black hole
accretion on galaxy evolution timescales \citep{raff11,mull12a,chen13},
in agreement with torque-limited growth.  Furthermore, \citet{hickox14}
have shown that accounting for short-term AGN variability may bring a wide
range of observations into agreement with an underlying SFR--$\dot{M}_{\rm
BH}$ correlation on cosmological timescales, including the observed weak
correlations between SFR and AGN luminosity in normal systems and general
trends in the observed AGN luminosity functions.   Hence, observations are
broadly consistent with the basic prediction of our torque-limited model
in that, when averaged over cosmological timescales, black hole accretion
rates track their host galaxies' SFRs.

\begin{figure} 
\begin{center}
\includegraphics[scale=0.5]{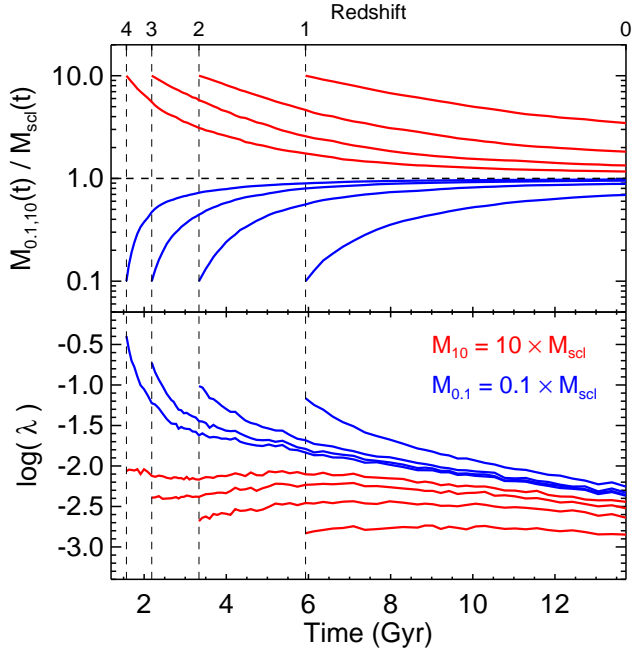} 
\end{center} 
\caption{{\it Top}: the impact of initial conditions on black hole growth.  For each host
galaxy, we consider black holes with initial masses that are either a factor
of 10 above ($M_{10}$; red) or below ($M_{0.1}$; blue) the corresponding
\Mbulge~relation and compare their evolution to that of a central black hole
initially consistent with the \Mbulge~relation ($M_{\rm scl}$).  Red and blue
solid lines show median values for the mass ratios $M_{10}/M_{\rm scl}$ and
$M_{0.1}/M_{\rm scl}$, respectively, for all host galaxies as a function of time.
Initial conditions are defined at a common redshift for all galaxies, which is
taken to be $z = $ 4, 3, 2, or 1, as indicated by the vertical dashed lines.
{\it Bottom}: evolution of accretion rates in Eddington units resulting
from the initial conditions defined in the top panel.  Red and blue solid
lines correspond to median values for black holes initially over-massive or
under-massive relative to the \Mbulge~relation at the starting redshift.}
\label{fig:conv} 
\end{figure}

\subsection{Convergence toward the Scaling Relations}\label{sec:conv}

Black holes tend to evolve onto the \Mbulge~relation corresponding to their
host galaxy regardless of the initial conditions \citep{ang13} and with no
need for mass averaging through mergers or additional self-regulation processes
(Section~\ref{sec:msig}).  Besides providing a natural explanation for the
observed scaling relations, this convergent behavior of gravitational torque
accretion may have significant implications for the accretion histories of
massive black holes and the interpretation of observations.

\begin{figure*} 
\begin{center}
\includegraphics[scale=0.8]{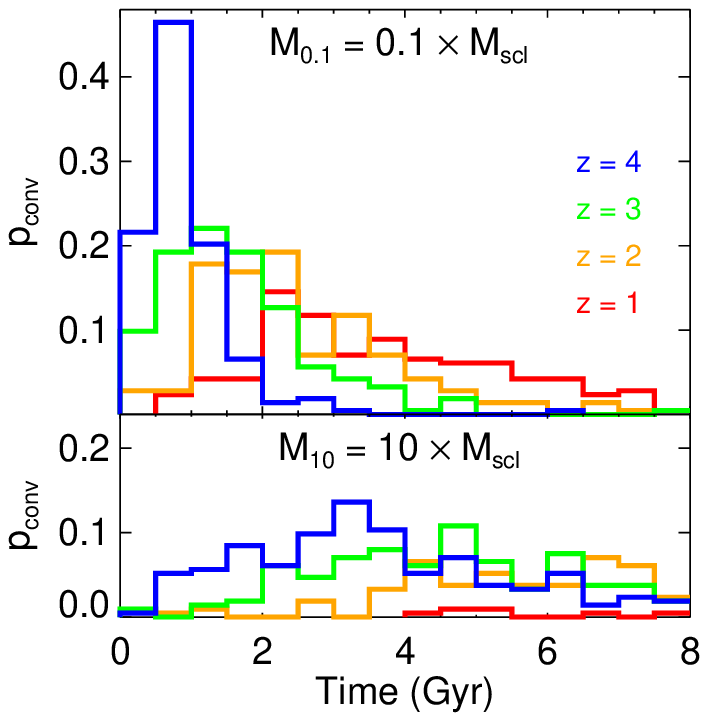}
\includegraphics[scale=0.8]{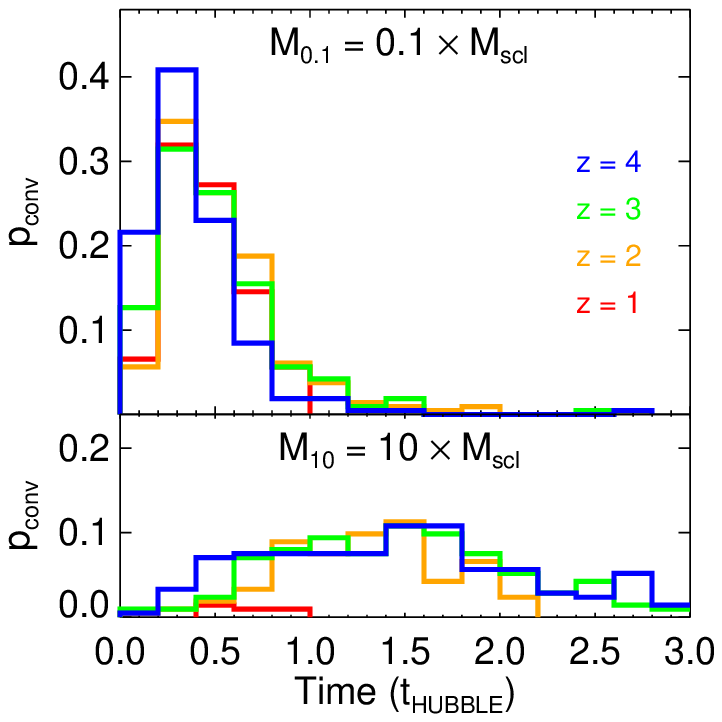} 
\end{center}
\caption{Probability for initially under-massive black holes ({\it top})
or initially over-massive black holes ({\it bottom}) of converging toward
the \Mbulge~relation in a given timescale.  The convergence timescale is
defined as the time required for a black hole with initial mass 10 times
above or below the \Mbulge~relation to grow to less than a factor of two
difference relative to a black hole that had an initial mass consistent
with the \Mbulge~relation at the starting redshift.  Blue, green, orange,
and red histograms show the probability distributions for initial conditions
defined at $z = $ 4, 3, 2, and 1, respectively (as in Figure~\ref{fig:conv}).
Convergence timescales are expressed in units of 1\,Gyr (left) or in units of
the Hubble time at each redshift (right).} 
\label{fig:conv_hist} 
\end{figure*}

Figure~\ref{fig:conv} provides further insight into this by comparing the
growth of central black holes with different initial masses under the evolution
of the same host galaxy.  For each of the \Ngal~simulated galaxies from our
primary sample, we follow the evolution of three black holes with an
initial mass
taken to be (1) consistent with the corresponding \Mbulge~relation at
$z = 4$, $M_{\rm scl}$, (2) a factor of 10 above, $M_{10} \equiv 10
\times M_{\rm scl}$, and (3) a factor of 10 below, $M_{0.1} \equiv
0.1 \times M_{\rm scl}$.  We then calculate the median value of the mass
ratios $M_{10}(t)/M_{\rm scl}(t)$ and $M_{0.1}(t)/M_{\rm scl}(t)$ over all
host galaxies as a function of time, which are shown by the red and blue
solid lines in the top panel of Figure~\ref{fig:conv}.  The same process
is repeated for starting redshifts $z = $ 4, 3, 2, and 1, where all host
galaxies are ``seeded" at the same redshift using black holes with initial
masses as defined above.

As expected from the middle panel of Figure~\ref{fig:Mbulge}, the initial
conditions for black hole growth are smoothed out by subsequent evolution,
resulting in mass ratios $M_{10}(t)/M_{\rm scl}(t)$ and $M_{0.1}(t)/M_{\rm
scl}(t)$ that approaches one with time.  Figure~\ref{fig:conv} (top panel)
allows us to infer the timescale in which torque-limited growth erases the
initial conditions and its dependence on redshift.  We find that over-massive
black holes require longer convergence timescales relative to black holes
with initial mass below the scaling relation.  Furthermore, the timescale
for convergence toward the \Mbulge~relation significantly increases with
decreasing starting redshift.  This is seen for initial black holes both
above and below the scaling relation.

This numerical experiment allows us to look at the effects of initial
conditions on the evolution of Eddington ratios.  The bottom panel of
Figure~\ref{fig:conv} shows the evolution of the median Eddington ratios
corresponding to the populations of black holes initially over-massive
or under-massive at different starting redshifts, as defined for the top
panel.  Given the dependence of gravitational torque rates on black hole
mass ($\lambda \propto M_{\rm BH}^{-5/6}$), under-massive black holes are
characterized by higher Eddington ratios relative to black holes lying on
the \Mbulge~relation.  Increased Eddington ratios only last for a period of
time given by the convergence timescale and, therefore, the evolution of
$\lambda$ is characterized by a rapid decrease at early times followed by
the usual decline at lower redshifts, as seen in Figure~\ref{fig:EddRatio}.
Similar arguments can be made for a population of over-massive black holes at
any given redshift.  In this case, Eddington ratios are strongly suppressed
initially and may even slightly increase with time if the mass decline relative
to the scaling relation supersedes the overall decline in Eddington ratios.
The net effect of having a population of over-massive black holes relative to
the \Mbulge~relation at any given redshift is a weaker evolution of $\lambda$
with time.

Figure~\ref{fig:conv_hist} shows quantitative predictions of the timescale
for convergence toward the \Mbulge~relation, which we define here as the
time required for a black hole with initial mass either 10 times above or
below to that corresponding to the \Mbulge~relation to grow to less than a
factor of two difference relative to a black hole that had an initial mass
consistent with the \Mbulge~relation at the starting redshift.  We compute
black hole convergence probabilities as a function of time after seeding
based on the number of host galaxies for which their central black holes did
converge in a given timescale.  As in Figure~\ref{fig:conv}, we take $z =
$ 4, 3, 2, and 1 as the starting redshifts.  The timescales are expressed in
Gyr for the left panel and scaled by the Hubble time corresponding to each
starting redshift in the right panel.

The convergence time probability distribution for under-massive black holes 
peaks at significantly shorter timescales relative to over-massive black holes
(Figure~\ref{fig:conv_hist}, left panel).  For example, the median
convergence timescale for under-massive black holes starting at $z = 4$
is $\sim 0.7$\,Gyr whereas for over-massive black holes it increases up to
$\sim 3.6$\,Gyr.  This is not unexpected, since the amount of mass required
to balance out the initial mass difference relative to the baseline mass
from the \Mbulge~relation is about 10 times higher for over-massive black
holes according to the definition adopted here.  Indeed, only $\sim 5$\,\%
of the over-massive black holes starting at $z = 1$ had enough time to
converge before the end of the simulation at $z = 0$, while $\sim 86$\,\%
of the under-massive black holes starting at $z = 1$ have converged.

Given some initial log-normal scatter, the mass-dependence of the convergence
timescales may produce a bias toward higher-mass black holes at later
times, since it takes longer for higher-mass black holes to evolve toward
the \Mbulge~relation relative to lower-mass black holes.  If the intrinsic
scatter of the \Mbulge~relation is higher at early times, this might
imply an increasing number of over-massive black holes at higher redshifts
that could be observed prior to convergence, as some observations suggest
\citep{treu07,decarli10,greene10,merloni10,bennert11,targett12}.  The initial
conditions as well as the redshift dependence of the 
convergence timescales may
thus have implications on the observed evolution of black hole mass to host
galaxy mass ratios.

Convergence probability distributions are indeed a strong function of
starting redshift.  The median convergence timescale and the spread of the
distribution both increase with decreasing redshift.  For under-massive black
holes, the characteristic (median) timescale increases from $\sim 0.7$\,Gyr
for $z = 4$ to $\sim 3.5$\,Gyr for $z = 1$, and the standard deviation of
the distribution increases by a factor $\sim 2.6$ with decreasing starting
redshift from $z = 4\rightarrow 1$.  Similar trends can be seen for the
distributions corresponding to over-massive black holes, despite not being
appropriately characterized at the lower redshifts given the fraction
of black holes for which convergence timescales are not well defined.
Interestingly, if we express convergence timescales in units of the Hubble
time for each starting redshift, $t_{\rm Hubble}(z)$, the resulting probability
distributions are remarkably similar (Figure~\ref{fig:conv_hist}, right panel).
The characteristic timescale for convergence toward the \Mbulge~relation
is $\sim 0.5 \times t_{\rm Hubble}$ and $\sim 1.5 \times t_{\rm Hubble}$
for under-massive and over-massive black holes, respectively, regardless of
the starting redshift.  This suggests that cosmological gas infall is the
ultimate physical mechanism driving the global evolution of massive black
holes and galaxies.

What is making black holes converge toward a similar mass regardless of the
initial conditions?  Let us consider a generic model in which the accretion
rate depends on the black hole mass with some power index $p$, $\dot{M}_{\rm
BH} = D(t) \times M_{\rm BH}^{p}$, where $D(t)$ contains all the explicit
dependencies on the host galaxy properties.  Let us now consider the growth of
two seed black holes with masses $M_{\rm a}(t)$ and $M_{\rm b}(t)$ evolving
at the center of an identical host galaxy.  The evolution of their mass ratio
is simply given by: 
\begin{equation}\label{eq:conv} 
\frac{d}{dt} \left ( \frac{M_{\rm a}}{M_{\rm b}} \right ) = D(t)
\, \frac{M_{\rm a}^{p}}{M_{\rm b}} \left [ 1 - \left ( \frac{M_{\rm a}}{M_{\rm b}} \right )^{1-p} \right ], 
\end{equation} 
where we have used $\dot{M}_{\rm
a} = D(t)\,M_{\rm a}^{p}$, $\dot{M}_{\rm b} = D(t)\,M_{\rm b}^{p}$, and the
fact the both black holes evolve under the same physical conditions $D(t)$.
Therefore, if $p < 1$, as is the case for the gravitational torque model
(Equation~(\ref{eq:torque})), the mass ratio $m_{\rm ab} \equiv M_{\rm a} /
M_{\rm b}$ will tend to approach one regardless of the initial conditions:
\begin{itemize} \item $dm_{\rm ab}/dt < 0$, \; if $m_{\rm ab} > 1$ \item
$dm_{\rm ab}/dt > 0$, \; if $m_{\rm ab} < 1$ \end{itemize}

Note that the exact opposite result applies to accretion models with $p >
1$, including the popular Bondi--Hoyle--Littleton parametrization \citep[$p
= 2$;][]{hoy39,bon44,bon52} and direct free-fall accretion \citep[$p =
2$;][]{hobbs12}.  Other examples of accretion parametrizations with $p < 1$
include the local viscous accretion rate \citep{debuhr11} and the radiation
drag model \citep{okamoto08}, neither of which have an explicit dependence
on black hole mass ($p = 0$).

The power index $p$ determines whether the initial conditions for black
hole growth, i.e. the initial black hole mass, tend to be erased ($p <
1$) or accentuated ($p > 1$) with subsequent evolution.  The timescale
for which black holes with different masses converge toward a similar
value depends on the initial mass ratio and the physical conditions $D(t)$
in the host galaxy.  Thus, the spread of the probability distributions in
Figure~\ref{fig:conv_hist} reflect the diversity of accretion histories for
our sample of host galaxies.  Note, however, that $p < 1$ alone does not
imply convergence toward the \Mbulge~relation specifically; the slope and
normalization is a non-trivial consequence of the physics included in the
black hole accretion parametrization.

Equation~(\ref{eq:conv}) implies that fine tuning of initial conditions may
be required if the main physical mechanism responsible for black hole growth
satisfies $p > 1$, since slightly different initial conditions could result in
rather different black hole masses at later times.  Black hole accretion rates
are defined to be positive and, therefore, any valid accretion parametrization
must satisfy $D(t) > 0$ at all times.  Thus, the only way to make a black hole
model with $p > 1$ less sensitive to the initial conditions is by introducing
some additional dependence on black hole mass that cannot be absorbed into
the power law dependence.  In practice, this is accomplished by having $D(t)$
depend on the accretion rate itself, $\dot{M}_{\rm BH} = D(t,\dot{M}_{\rm BH})
\times M_{\rm BH}^{p}$, which is usually justified in self-regulated models
by assuming that feedback from the accretion process has a direct effect
on the accretion flow itself.  This simple argument shows why a nonlinear
feedback loop is required to regulate black hole growth when using an accretion
parametrization with a strong dependence on black hole mass \citep{ang13},
and why the torque-limited model does not require explicit self-regulation.

\begin{figure} 
\begin{center}
\includegraphics[scale=0.4]{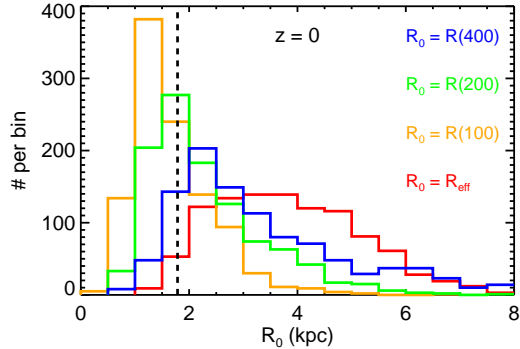} 
\end{center}
\caption{Distribution of radial apertures ($R_{0}$) at $z = 0$
corresponding to different definitions, including the smallest radius enclosing (1)
100 (orange), (2) 200 (green), or (3) 400 (blue) gas and star
particles, and (4) the effective radius of the host galaxy, $R_{\rm
eff}$ (red).  All galaxies containing at least 400 gas particles and 400
star particles at $z = 0$ have been included.  
The vertical dashed line indicates the gravitational softening length of the simulation in physical kpc at $z = 0$.} 
\label{fig:R0} 
\end{figure}

\begin{figure} 
\begin{center}
\includegraphics[scale=0.95]{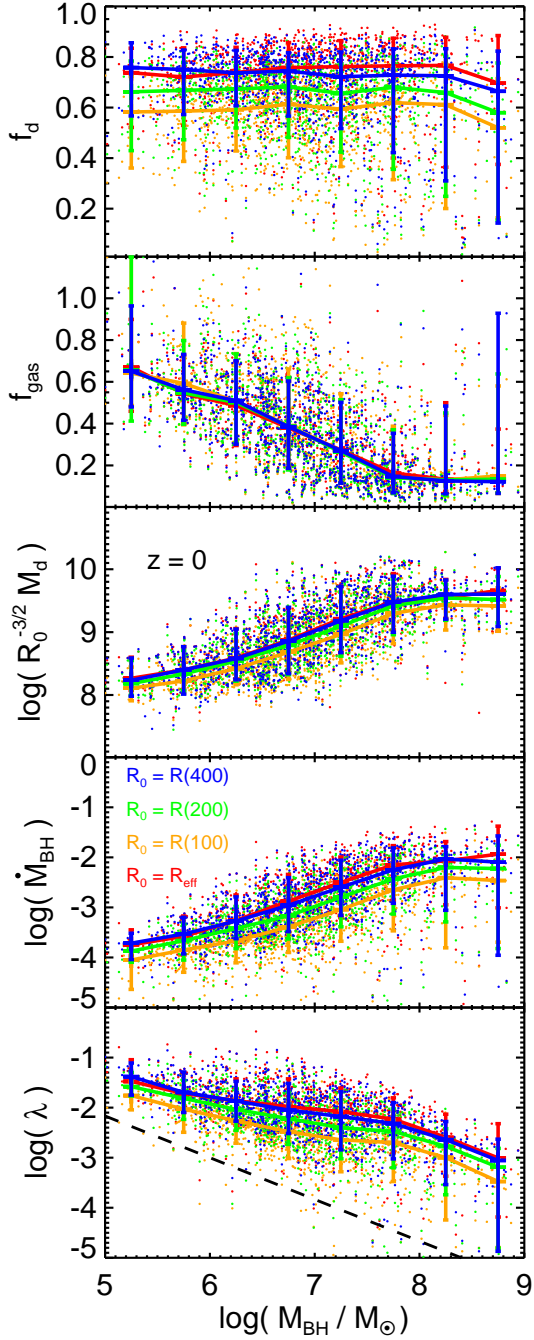} 
\end{center} 
\caption{Effects
of different radial apertures ($R_{0}$) on the inferred gravitational torque
rates for all galaxies with at least 400 gas and star particles at $z = 0$.
Accretion rates are calculated for different definitions of the radial aperture
$R_{0}$, including the smallest radius enclosing (1) 100 (orange),
(2) 200 (green), or (3) 400 (blue) gas and star particles,
and (4) the effective radius of the host galaxy, $R_{\rm eff}$ (red).
We show, from top to bottom, some of the key physical quantities defined
in Equation~(\ref{eq:torque}) as a function of black hole mass: $f_{\rm d}$,
$f_{\rm gas}$, and $R_{0}^{-3/2}M_{\rm d}$ (in units of kpc$^{-3/2}$\,\Msun),
as well as the resulting accretion rates (in units of \Msun\,yr$^{-1}$)
and Eddington ratios, where we have assumed that black holes lie on the
\Mbulge~relation with 0.5 dex scatter in black hole mass.  Individual black
holes are represented by points with the colors corresponding to the different
radial apertures.  Points with error bars connected by solid lines show median values
within logarithmically spaced bins in $M_{\rm BH}$ and the central 75\,\% of
the distribution.  The black dashed line in the bottom panel corresponds to
the scaling $\lambda \propto M_{\rm BH}^{-5/6}$.} 
\label{fig:res} 
\end{figure}

\section{Numerical robustness}\label{sec:res}

In \citet{ang13}, we conducted an extensive analysis of how the inferred
gravitational torque rates and the resulting black hole--galaxy scaling
relations depend on a variety of parameters and implementation details
including: numerical resolution, the mass retention rate ($\epsilon_{\rm m}$),
and different stellar feedback models.  In this section, we complement our
previous study by presenting additional tests, focusing primarily on the
convergence of results for different radial apertures ($R_{0}$) and on the
impact of our quenching prescription on the co-evolution of black holes
and galaxies.
Specific numerical convergence tests for our bulge--disk decomposition procedure are presented in the Appendix, where a direct comparison is made to other decomposition methods commonly used in the literature.

\subsection{Evaluating Gravitational Torque Rates}\label{sec:resT}

We select all galaxies containing at least 400 gas particles and 400 star
particles at $z = 0$, so that different radial apertures can be defined for
a common galaxy sample.  To evaluate the effects of $R_{0}$ on the
inferred gravitational torque rates, we define radial apertures that enclose
at least either (1) 100 ($R_{100}$), (2) 200 ($R_{200}$), or
(3) 400 ($R_{400}$) gas and star particles, or, alternatively, we use
(4) the effective radius of the host galaxy, $R_{0} = R_{\rm eff}$,
as in Section~\ref{sec:sfagn}.  Figure~\ref{fig:R0} shows the distribution
of radial apertures at $z = 0$ corresponding to the different definitions.

\begin{figure*} 
\begin{center}
\includegraphics[scale=0.33]{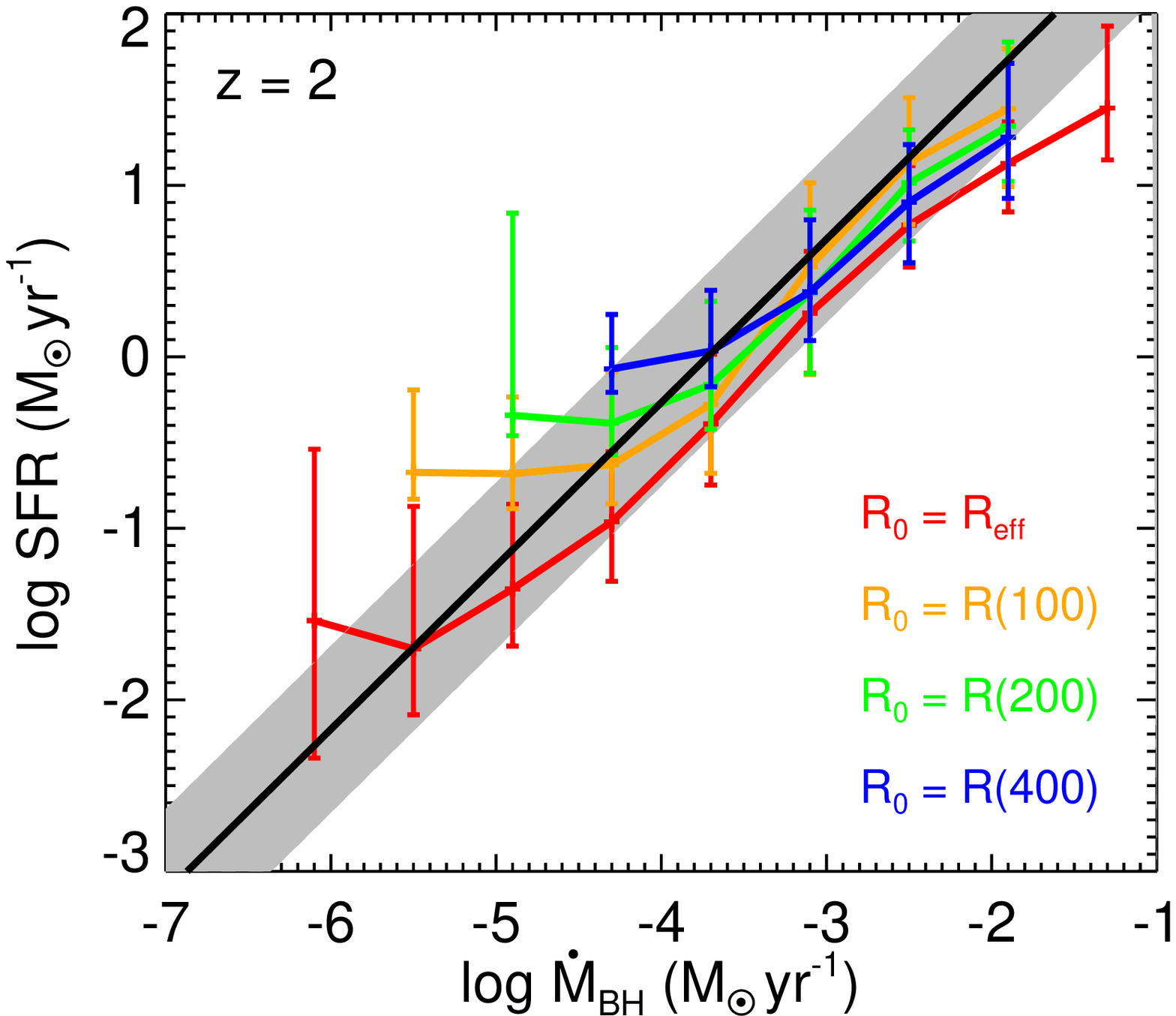}
\includegraphics[scale=0.33]{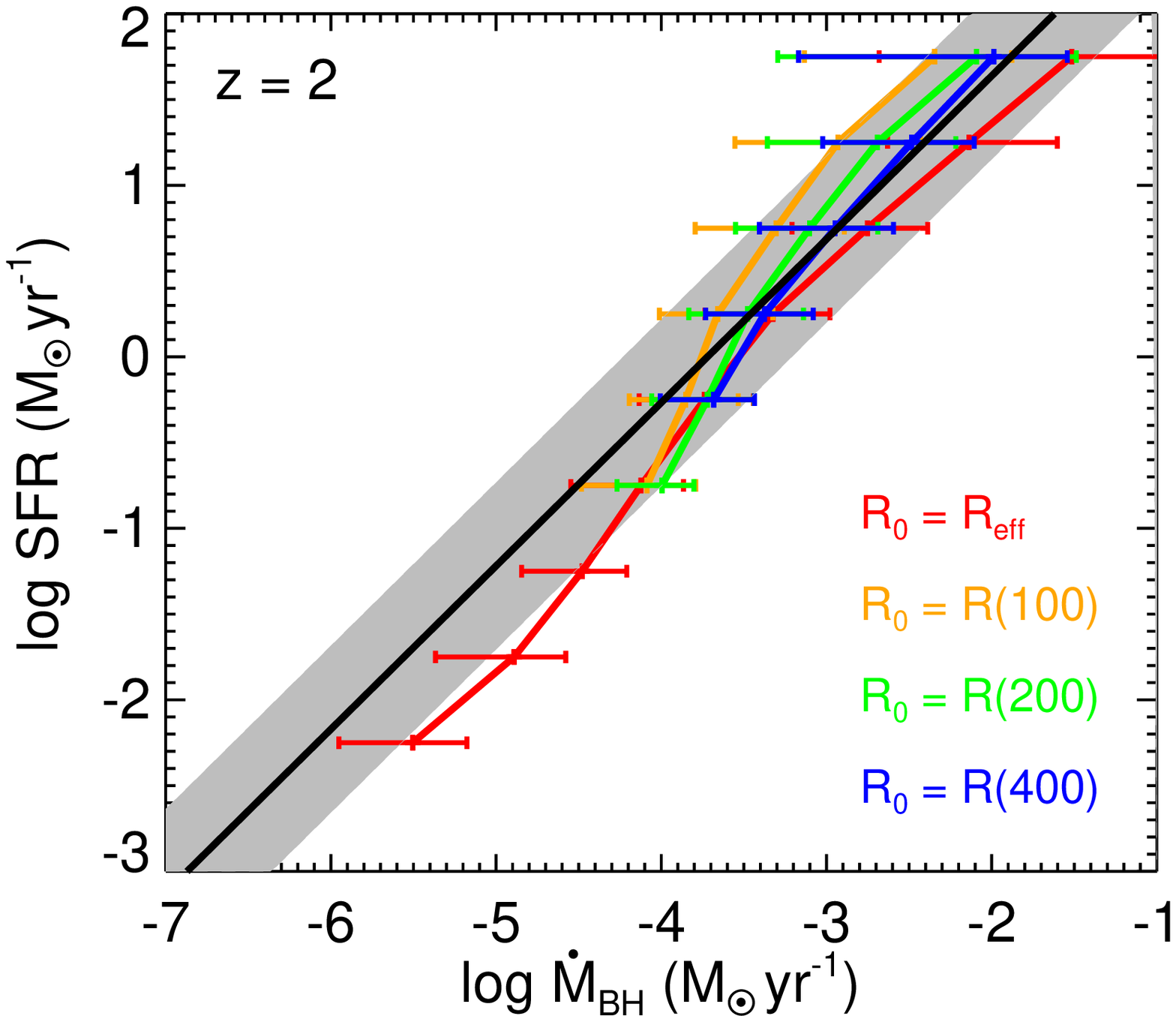}
\includegraphics[scale=0.33]{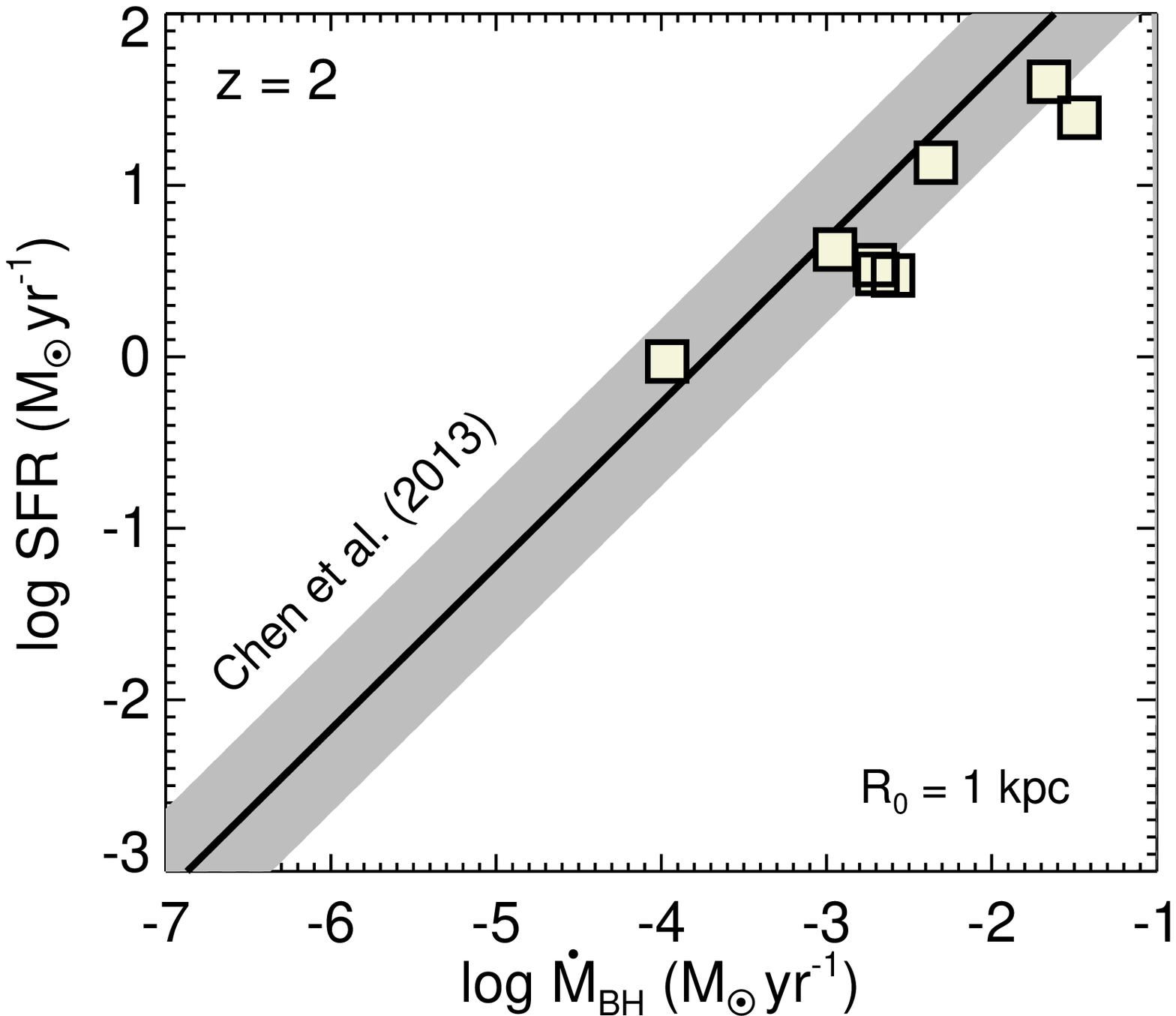} 
\end{center} 
\caption{Numerical convergence test for the SFR--$\dot{M}_{\rm BH}$ diagram at $z = 2$.
{\it Left}: median host galaxy SFRs within logarithmically spaced bins in
$\dot{M}_{\rm BH}$, where the error bars enclose 75\,\% of the distribution.
Accretion rates are calculated as in Figure~\ref{fig:SFRvsMdot} for different
definitions of the radial aperture $R_{0}$, including the smallest radius
enclosing (1) 100 (orange), (2) 200 (green), or (3) 400
(blue) gas and star particles, and (4) the effective radius of the
host galaxy, $R_{\rm eff}$ (red).  Here, we include all $z = 2$ galaxies with
the minimum number of gas and star particles required for each definition
of $R_{0}$.  {\it Middle}: same as the left panel for the median black
hole accretion rates within logarithmically spaced bins in the host galaxy SFR.
{\it Right}: filled squares show the SFR--$\dot{M}_{\rm BH}$ relation at $z
= 2$ for eight re-simulated galaxies from \citet{ang13}, where black hole
accretion rates were calculated using a fixed radial aperture $R_{0} = 1$\,kpc
(physical).  The mass resolution of these cosmological zoom simulations is
$\sim 20$ times higher relative to the full cosmological simulation employed
here. The black solid line and the gray shaded region in each panel correspond
to the SFR--$\dot{M}_{\rm BH}$ relation of \citet{chen13} and the estimated
uncertainty in the normalization.} 
\label{fig:SFRvsMdot_res} 
\end{figure*}

Figure~\ref{fig:res} shows some of the key galaxy properties entering into the
calculation of gravitational torque rates as a function of black hole mass
at $z = 0$.  In particular, we calculate the disk fraction ($f_{\rm d}$),
the gas fraction relative to the disk mass ($f_{\rm gas}$), and the total
disk mass within $R_{0}$ ($M_{\rm d}$) for the radial apertures defined above.
In addition, we evaluate the accretion rates as well as the Eddington ratios
corresponding to the different radial apertures by assuming that black holes
lie on the \Mbulge~relation with 0.5 dex scatter in black hole mass.

The top panel of Figure~\ref{fig:res} shows that there is significant scatter
in the disk fraction, with median $f_{\rm d}$ values roughly independent of black hole mass
independent of the definition of $R_{0}$.  
There is actually an increasing number of black holes with host galaxies having lower $f_{\rm d}$ values at higher masses, but this trend is not as clear as we would expect from observations in the local universe, where the majority of $\sim 10^9$\,\Msun~black holes reside in elliptical galaxies.
This is, however, not surprising given the low number of the most massive systems in our simulation as well as the definition of disk fraction employed here, which may include contributions from rotating bulges (Section~\ref{sec:accr}). 
Nonetheless, we find evidence for
a systematic trend of increasing disk fractions for larger radial apertures.
Median $f_{\rm d}$ values within logarithmically spaced bins in $M_{\rm
BH}$, indicated by points with error bars connected by solid lines, increase
by about 25\,\% for the radial aperture $R_{100}$ relative to $R_{400}$.
This might be expected given that, at the resolution of our cosmological
simulation, $R_{400}$ is comparable to the effective radius ($R_{\rm eff}$)
and galaxies are more likely bulge-dominated in their central regions.
Furthermore, the radial apertures $R_{100}$ and $R_{200}$ are comparable to or even smaller than the gravitational softening length of the simulation (Figure~\ref{fig:R0}).  The lack of gravitational resolution could result in the overestimation of bulge masses (and hence the underestimation of $f_{\rm d}$) owing to the increased random motions and the overall reduction in the amount of ordered circular motion for an artificially shallower gravitational potential \citep{ang14}. 
Note, however, that this is greatly compensated by our definition of $f_{\rm d}$ (formally an upper limit to the disk fraction), resulting in better resolution convergence relative to more restrictive bulge--disk decomposition methods (the Appendix).

Gas fractions relative to the disk mass are better converged with respect to $R_{0}$, as shown in
the second panel of Figure~\ref{fig:res} (from top to bottom).  The median
values of $f_{\rm gas}$ are essentially coincident for the different radial
apertures, clearly decreasing for higher mass black holes and, therefore,
higher mass galaxies, as expected \citep{dav11a}.  Note that $f_{\rm gas}$
here is defined as the ratio of the gas mass to the total disk mass and may thus be greater
than one (Equation~(\ref{eq:fgas})).

The disk mass normalized by the radial aperture, $R_{0}^{-3/2}M_{\rm
d}$, shows, again, some discrepancy between the different definitions
of $R_{0}$.  The radial aperture $R_{400}$ yields median values about 0.2
dex above those obtained with $R_{100}$.  Besides the obvious dependence on
$f_{\rm d}$, different radial density profiles may also affect the inferred
$R_{0}^{-3/2}M_{\rm d}$ values.  Indeed, the comparison between $R_{400}$
and $R_{200}$ suggests that we are approaching numerical convergence.
Interestingly, all radial apertures show a significant increase in median
$R_{0}^{-3/2}M_{\rm d}$ values with increasing black hole mass.  Given that
there is no clear trend for $f_{\rm d}$ with  $M_{\rm BH}$, this suggests
that more massive galaxies are more compact.

The bottom two panels of Figure~\ref{fig:res} show the accretion rates and
Eddington ratios as a function of $M_{\rm BH}$ resulting from the galaxy
properties described in the upper panels.  The combined effects of the radial
aperture result in median $\dot{M}_{\rm BH}$ and $\lambda$ values that may
vary by about a factor of two to three for the different definitions of $R_{0}$.
Despite this, all radial apertures considered here yield very similar trends
with $M_{\rm BH}$ and, therefore, a simple normalization factor may bring the
results into better agreement.  Note that the radial aperture must be chosen
as a trade off between (1) the number of resolution elements required
to characterize the morphology of the galaxy and (2) smaller physical
radial apertures that provide a better prediction for the gas inflows at
sub-parsec scales \citep{hop11}.  Thus, strict convergence with respect to
$R_{0}$ may only be reached by increasing the mass and force resolution of
the simulation \citep{ang13}.

Besides illustrating the robustness of our methodology to changes in
$R_{0}$ at fixed resolution, Figure~\ref{fig:res} allows us to gain further
insight into the dependence of Eddington ratios on black hole mass.
Indeed, the increased dynamic range relative to our baseline sample of
\Ngal~black holes (Figure~\ref{fig:LratiovsMbh}) reveals a clear trend
for $\lambda$ values to decrease with $M_{\rm BH}$.  Similar  results are
reproduced at all redshifts, with consistently weaker trends relative to
the expected $\lambda \propto M_{\rm BH}^{-5/6}$ dependence, as described
in Section~\ref{sec:Edd}.  Higher mass galaxies are more compact than lower
mass galaxies while having a similar disk fraction within $R_{0}$, resulting
in higher values of $R_{0}^{-3/2}M_{\rm d}$.  Moreover, Figure~\ref{fig:res}
shows that, at fixed redshift, accretion rates are primarily determined
by $R_{0}^{-3/2}M_{\rm d}$, yielding higher $\dot{M}_{\rm BH}$ values with
increasing $M_{\rm BH}$.  Note that the gas fraction decreases with increasing
$M_{\rm BH}$ but the torque model is only weakly dependent on $f_{\rm gas}$.
The resulting accretion rates, when normalized by black hole mass, yield
Eddington ratios decreasing with $M_{\rm BH}$ roughly as $\lambda \propto
M_{\rm BH}^{-1/2}$.  We do not pursue this further given the limited mass
range of the sample of our \Ngal~black hole sample for which their evolution
can be constrained self-consistently from $z \gtrsim 4$ down to $z =  0$.

\subsection{The SFR--$\dot{M}_{\rm BH}$ Relation}

We now turn back to the SFR--$\dot{M}_{\rm BH}$ diagram in
Figure~\ref{fig:SFRvsMdot_res}, where we evaluate the effects of different
radial apertures and numerical resolution on the inferred SFR--AGN connection
at $z = 2$.  As in Figure~\ref{fig:res}, we calculate accretion rates
using different values of $R_{0}$ and assuming that black holes lie on
the \Mbulge~relation with 0.5 dex scatter in black hole mass.  However,
we include here all $z = 2$ host galaxies with at least the minimum number
of particles sufficient to define each of the different radial apertures;
this results in a significantly larger dynamic range for $R_{0} = R_{\rm
eff}$ relative to, for example, $R_{0} = R_{200}$ since the effective radius
can be defined for all galaxies.

The left panel of Figure~\ref{fig:SFRvsMdot_res} shows the median SFRs
within logarithmically spaced bins in $\dot{M}_{\rm BH}$ corresponding to
the different radial apertures, with error bars indicating the 12.5 and
87.5 percentiles.  Besides the increased dynamic range, smaller apertures
yield higher SFRs for a given accretion rate, corresponding to higher mass
galaxies and higher mass black holes, as expected from our discussion
in Section~\ref{sec:resT}.  However, this is not the case in the low
$\dot{M}_{\rm BH}$ regime, where the SFR--$\dot{M}_{\rm BH}$ relation
flattens out and smaller radial apertures seem to result in lower SFRs for
a given $\dot{M}_{\rm BH}$.  This partly owes to selection effects, since
larger radial apertures can be defined only for higher mass galaxies with
correspondingly higher SFRs.

At a more fundamental level, one expects a flattening of the 
SFR--$\dot{M}_{\rm BH}$
relation at low $\dot{M}_{\rm BH}$ to occur when calculating median
SFRs as a function of $\dot{M}_{\rm BH}$ owing to the different characteristic
variability timescales of star formation and black hole accretion and because
black holes spend more time accreting below the mean value \citep{hickox14}.
This trend disappears when we invert the SFR--$\dot{M}_{\rm BH}$ relation
and calculate median black hole accretion rates within SFR bins, as shown
in the middle panel of Figure~\ref{fig:SFRvsMdot_res}.  In this case,
the gravitational torque model produces a SFR--$\dot{M}_{\rm BH}$ relation
similar to that of \citet{chen13} even in the low $\dot{M}_{\rm BH}$ regime
and for all radial apertures.  Note that the scatter increases for higher
$\dot{M}_{\rm BH}$ values owing to the small number of galaxies with high SFRs.

As an additional consistency check, the right panel of
Figure~\ref{fig:SFRvsMdot_res} shows the location in the SFR--$\dot{M}_{\rm
BH}$ plane of eight $z = 2$ re-simulated galaxies taken form \citet{ang13}.
The higher resolution of these cosmological zoom simulations (a 20 times
higher mass resolution relative to our primary cosmological simulation in this
work) allowed us to compute galaxy properties within a fixed radial aperture
$R_{0} = 1$\,kpc (physical) for all galaxies at all times.  Encouragingly,
the inferred central black holes occupy the region of the diagram expected
for their host galaxy SFRs, in agreement with our current results and
independent of the differences in resolution, the operational definition of
$R_{0}$, and even feedback effects, since these simulations did not include
any quenching mechanism.

\begin{figure} 
\begin{center}
\includegraphics[scale=0.45]{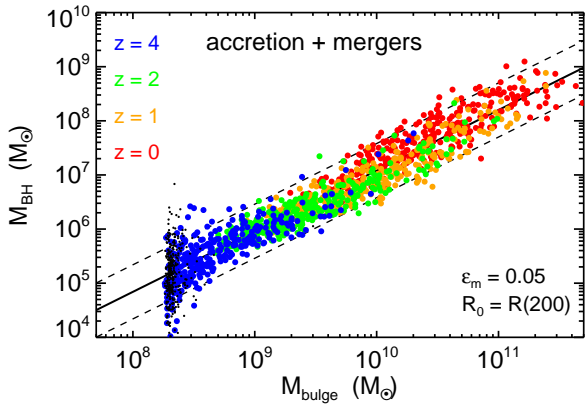} 
\end{center}
\caption{\Mbulge~relation at $z = $ 0 (red), 1 (orange), 2 (green), and 4
(blue) for black holes growing through torque-limited accretion and mergers.
Host galaxies are taken from a $2 \times 512^3$ cosmological simulation in a
$[32\,\hmpc]^3$ comoving box including no high mass galaxy
quenching mechanism.  Masses of black
hole seeds (shown as small black dots) and merging black holes are randomly
selected from a log-normal distribution corresponding to the \Mbulge~relation
for the appropriate galaxy and time step, assuming a 0.5 dex scatter in black
hole mass. The black solid line shows the \Mbulge~relation of \citet{har04}
while the black dashed lines indicate a 0.5 dex scatter in black hole mass.}
\label{fig:noQ} 
\end{figure}

 \subsection{Effects of Star-formation Quenching}

Our primary cosmological simulation incorporates a heuristic prescription
for star-formation quenching that is tuned to reproduce the observed
exponential cutoff in the high-mass end of the stellar mass function at $z =
0$ \citep{dav13}.  Given that AGN feedback is currently the best candidate
for suppressing star formation in the high-mass regime, it is important to
evaluate the impact of our quenching prescription on the inferred co-evolution
of black holes and galaxies.  To do so, we have repeated all calculations for
an additional cosmological simulation with the same size and resolution but
including no quenching mechanism.  Specifically, this is the r32n512vzw run
described in \citet{dav13}.  We note that this simulation adopts a slightly
different model for galactic outflows, but the differences are confined
to galaxies with velocity dispersions $\sigma<75$~km~s$^{-1}$, whereas the
galaxies we consider here are generally larger.  Hence the main difference
for black hole growth should reflect the differences owing to our heuristic
quenching prescription.

Figure~\ref{fig:noQ} shows the \Mbulge~relation obtained at different redshifts
when we include no quenching prescription.  Here, we have applied the same
selection criteria as in Section~\ref{sec:msig}, including only galaxies
that contain at least 200 gas and 200 star particles at all times and that
are identified in the cosmological simulation as early as $z \geq 4$.
This results in a higher number of galaxies extending to higher masses
relative to the simulation with star-formation quenching.  Despite the
strong difference at the high-mass end of the stellar mass function, our
no-quenching simulation yields black holes and host galaxies evolving on
average along the \Mbulge~relation, just as in our quenching simulation
(see the bottom panel of Figure~\ref{fig:Mbulge}), and for the same mass
retention rate $\epsilon_{\rm m} = 0.05$.    
Interpreting our heuristic quenching prescription as a plausible effect of AGN feedback acting on the host galaxy, this would suggest that AGN feedback is not driving the connection between black holes and galaxies even if it is responsible for star-formation quenching.
All other properties of torque-limited
growth analyzed here such as the convergence toward the scaling relations
and the characteristic distribution and evolution of Eddington ratios are
reproduced by our no-quenching simulation, providing further support for
the numerical robustness of our methodology.

Overall, our results are mildly sensitive to the radial aperture considered,
in the sense that smaller apertures result in lower accretion rates.
However, the overall trends are unchanged, hence a modest re-normalization
of $\epsilon_m$ could compensate for the differences, and our general
conclusions are unaffected by this choice.  Even though our heuristic
quenching prescription results in significantly fewer massive galaxies,
it has little impact on the black hole properties at a given bulge mass.

\section{Discussion and conclusions}\label{sec:dis}

In this paper, we applied and extended the methodology developed in
\citet{ang13} to infer the role of accretion driven by gravitational torques
on the evolution of massive black holes at the centers of star-forming galaxies
over cosmic time.  By combining the analytic model of \citet{hop11} with full
cosmological hydrodynamics simulations, we have (1) constrained the
physics driving the observed black hole--galaxy scaling relations, (2)
evaluated the relative importance of black hole mergers for the total growth,
(3) presented predictions for the distribution and evolution of
Eddington ratios, and (4) investigated the global connection between
black hole accretion and star formation.  Our main results can be summarized
as follows:

\begin{enumerate}

\item Torque-limited growth yields black holes and host galaxies evolving
on average along the \Mbulge~relation from early times down to $z = 0$.
The normalization of the scaling relation depends on the mass retention
rate $\epsilon_{\rm m}$, which represents the fraction of the inflowing gas
feeding the accretion disk from galactic scales that is finally accreted
by the central black hole.  We find that $\epsilon _{\rm m}\approx 5$\,\%
provides an appropriate normalization, implying that $\sim 95$\,\% of the
mass inflow at sub-parsec scales does not make it onto the black hole and
may be lost to winds and outflows.

\item By identifying galaxy mergers down to mass ratios of 1:5, we find
that smooth accretion represents most of black hole growth and dominates
the overall evolution in the \Mbulge~plane.  Black hole mergers represent
typically a small fraction of the total growth except in some exceptional
cases with numerous mergers happening preferentially at low redshift.

\item Black holes tend to evolve onto the \Mbulge~relation corresponding
to their host galaxy independent of the initial conditions and with no need
for mass averaging through mergers or additional self-regulation processes.
The characteristic convergence time scale for black holes starting a factor of
ten above or below the \Mbulge~relation is about 0.5 and 1.5 times the Hubble
time for initially under-massive and over-massive black holes, respectively.

\item The weak dependence of gravitational torque rates on black hole
mass plays a key role in the overall convergence behavior.  For accretion
parametrizations of type $\dot{M}_{\rm BH} \propto M_{\rm BH}^p$, it can be
shown that the power index $p$ determines whether the initial conditions
for black hole growth tend to be erased ($p < 1$), as is the case for
the gravitational torque model ($p = 1/6$), or accentuated ($p > 1$)
with subsequent evolution.  This implies the need for additional feedback
self-regulation for accretion models strongly dependent on $M_{\rm BH}$,
such as the popular Bondi parametrization ($p = 2$).

\item Eddington ratios averaged over galaxy evolution timescales can be
described at all redshifts by a broad log-normal distribution with a median
value evolving roughly as $\lambda_{\rm MS} \propto (1+z)^{1.9}$, suggesting
the existence of a main sequence for AGN analogous to the cosmic evolution of
specific SFRs.  Torque-limited accretion yields typical average Eddington
ratios $\lambda > 10$\,\% at early times, a smooth average transition
between radiatively efficient to radiatively inefficient accretion modes at
$z \approx 1$--2, and typical present day values of $\lambda \lesssim 1$\,\%.

\item The width of the distribution of Eddington ratios increases at lower
redshifts, with an extended tail toward low luminosities in radiatively
inefficient accretion.  The combined dependence of gravitational torque
rates on black hole mass and gas+stellar surface density yields a weak trend
of decreasing median Eddington ratios with increasing black hole mass at
all redshifts.

\item Torque-limited growth predicts a connection between SFR and AGN activity
on timescales comparable to the dynamical timescale of galaxies, resulting
in a close-to-linear $\langle \dot{M}_{\rm BH} \rangle \propto$ SFR relation
for the average black hole accretion rate.  Cosmological gas infall and
transport of angular momentum in the galaxy by gravitational instabilities
appear to be the primary drivers of this relation.  The SFR--AGN connection
can have large fluctuations on a galaxy-by-galaxy basis given the strong
variability of black hole accretion at all timescales.

\end{enumerate}

Encouragingly, our main results are robust to changes in a variety of
implementation details.  Despite the apparently complicated form of
the gravitational torque model, by comparing to our previous zoom runs,
its application to simulations of different resolution yields reasonably
good numerical convergence.  We have identified a slight trend for higher
accretion rates with increasing radial aperture $R_{0}$ at fixed resolution,
suggesting that higher resolution simulations are needed when $R_{0}$
becomes comparable to the effective radius of the galaxy.  Nonetheless,
the uncertainty in $R_{0}$ described here seems reasonable compared to the
typical uncertainties of self-regulated accretion models on, e.g.,
the frequency at which feedback events occur or the number
of resolution elements over which feedback energy or momentum is injected
\citep[e.g.,][]{booth09,dubois12,choi12,newton13}.  
At a sub-grid level, it is assumed that cooling is efficient and most of the gas forms a rotationally supported disk, but the gravitational torque model does not explicitly depend on the resolved gas thermodynamics \citep{hop11}, which is a major source of uncertainty in galaxy formation simulations.
In addition, torque-limited growth yields roughly similar results regardless of stellar feedback effects
\citep{ang13}.  
Different parametrizations of star formation (via an
effective equation of state) or stellar feedback may yield rather different
structural properties of galaxies \citep{ang14}, but the scalings arising
for the response of gas+stellar systems to gravitational torques remain
the same: black holes and galaxies simply move along the scaling relations
without changing the overall behavior of the gravitational torque model.

Our combined results imply a fuelling-controlled scenario in which black
hole growth is primarily governed by the amount of gas supply from galactic
scales and not by the direct interaction of feedback energy or momentum from
the accretion flow with the surrounding galaxy ISM \citep{escala06,escala07}.
AGN feedback in the form of winds and outflows from the accretion flow may have
a strong impact on the host galaxy and actually represents a significant mass
loss relative to the available gas supply, thereby strongly suppressing black
hole growth.  However, it is the rate at which non-axisymmetric perturbations
to the stellar potential drive gas into shocks that dissipate energy and
angular momentum that determines the overall rate of black hole growth and the
connection between star formation and AGN activity on cosmological timescales.

There remain, however, significant uncertainties that require further
investigation.  A key assumption in this work is that only a small fraction of
the gas feeding the accretion disk from galactic scales is finally accreted by
the black hole.  We have parametrized the mass lost to winds and outflows in
the accretion disk with the simplest possible model, by assuming a constant
average mass retention rate ($\epsilon_{\rm m}$) for all black holes at all
times.  Its value, $\epsilon_{\rm m} = 0.05$, has been determined by requiring
consistency with the \Mbulge~relation of \citet{har04}, but it is subject to
uncertainties including a degeneracy with the nuclear star formation law or the
exact definition of the \Mbulge~relation \citep{hop11,ang13}.  Despite this
simplistic assumption, the inferred average mass retention rate seems roughly
consistent with a number of observational studies and theoretical expectations.

Outflows appear to be a common feature of geometrically thick accretion
disks, usually ascribed to radiatively inefficient flows forming at very
low Eddington ratios \citep[$\lambda \lesssim 0.01$;][]{narayan08}, or
``slim disks" forming at super-critical rates \citep[$\lambda \gtrsim
1$;][]{abramowicz88}.  A wide range of simulations of hot accretion flows
show strong outflows that may carry away a significant fraction of the mass
inflow rate \citep{yuan12,bu13,li13,sadowski13}.  The radial profile of the
mass outflow rate can be described as a power-law in radius, $\dot{M}_{\rm
wind}(r) \propto r^{s}$, with typical power index values $s \sim 0.4$--1
\citep{yuan12}.  Thus, the black hole accretion rate can be calculated in
terms of the torque-limited inflow rate as: 
\begin{equation} 
%\dot{M}_{\rm BH} = \left [ \, 1 + \left ( \frac{r_{\rm out}}{r_{\rm in}} \right )^{s} \, \right
%]^{-1} \dot{M}_{\rm Torque}(r_{\rm out}), 
\dot{M}_{\rm BH} = \left [ \, 1 + \left ( \frac{r_{\rm out}}{r_{\rm in}} \right )^{s} \, \right
]^{-1} \dot{M}_{\rm Torque},
\end{equation} 
where $r_{\rm in}$ and $r_{\rm out}$ represent the inner and outer radii of the region where
accretion-driven outflows are launched and $\dot{M}_{\rm Torque}$ represents the gas inflow rate at $r_{\rm out}$ driven by gravitational instabilities at larger scales.  
If we take $s = 0.5$ \citep[e.g.][]{bu13} and assume
that the radial extent of the accretion disk powering outflows spans between
two and four orders of magnitude relative to the inner radius, $r_{\rm out}
/ r_{\rm in} \approx 10^{2}$--$10^{4}$ \citep{tombesi12}, we obtain a range
for the effective mass retention rate $\epsilon_{\rm m} \approx 1$--10\,\%,
roughly consistent with our inferred value.  It is less clear from simulations
whether radiatively efficient, geometrically thin disks \citep{shakura73}
produce outflows with similar scalings \citep{li13,sadowski13}, but significant
mass loss through magnetically or radiation-driven winds may also occur
\citep{proga08,ohsuga11}; nonetheless, there is substantial observational
evidence for strong outflows in the ``quasar mode" feedback, corresponding
to radiatively efficient accretion \citep{fabian12}.

It is illustrative to obtain a rough estimate of the average kinetic power
implied in our model.  The kinetic luminosity can be defined in terms of
the kinetic efficiency $\epsilon_{\rm k}$: \begin{equation} L_{\rm k} \,
= \, \frac{1}{2} \dot{M}_{\rm wind} v^{2} \,  \equiv  \, \epsilon_{\rm k}
\dot{M}_{\rm BH} c^{2}, \end{equation} \begin{equation} \epsilon_{\rm k}
\, = \, \frac{1}{2} \left ( \frac{v}{c} \right )^{2} \left ( \frac{1 -
\epsilon_{\rm m}}{\epsilon_{\rm m}} \right ), \end{equation} where $v$
is the outflow velocity and the definition of $\epsilon_{\rm m}$ implies
$\dot{M}_{\rm BH} = \dot{M}_{\rm Torque} - \dot{M}_{\rm wind}$.  Thus, for
mass-weighted outflow velocities in the range $10^{3}$--$10^{4}$\,km\,s$^{-1}$,
we obtain a plausible range for the kinetic efficiency $\epsilon_{\rm k}
\approx 10^{-4}$--$10^{-2}$, or between 0.1--10\,\% of the bolometric
luminosity assuming $\eta = 0.1$.  For comparison, the AGN synthesis model
of \citet{merloni08} yields a total integrated average kinetic efficiency
$\epsilon_{\rm k} \approx 3$--$5 \times 10^{-3}$, and 
self-regulated models
of black hole growth in galaxy-scale simulations require $\sim 0.5$--15\,\%
of the bolometric luminosity to be injected into the surrounding medium in
order to affect the host galaxy and possibly reproduce the observed black hole--galaxy
scaling relations \citep{dimatteo05,hopelv10,dubois12}.

Observations show that winds and outflows are ubiquitous in AGN for a wide
range of host galaxy properties \citep{reynolds97,veilleux05,fabian12},
though direct constraints on mass outflow rates have proven difficult
to obtain.  \citet{tombesi12} find mass loss rates $\dot{M}_{\rm wind} /
\dot{M}_{\rm BH} \sim 0.05$ for ultrafast outflows in radio-quiet AGNs,
significantly lower than the mass loss required by torque-limited growth,
while a comparison between kinetic wind luminosity to bolometric luminosity
for the sample of AGN analyzed by \citet{king13} suggests typical ratios of
mass loss in winds to black hole accretion comparable to our expectations.
The inferred outflowing masses and velocities likely depend on the distance
from the black hole to the wind material \citep{tombesi13} and the amount
of entrainment of surrounding material.  Indeed, the total mass-loss in
galaxy-scale AGN-driven outflows may exceed the SFR of the entire galaxy
\citep{feruglio10,rupke11,sturm11,maiolino12}.

At present, it is difficult to assess
how much mass is lost in outflows relative
to the gas inflowing at a given radius, but future observations and numerical
simulations will provide tighter constraints.  Despite the complexity inherent
to black hole accretion flows and outflows, it is encouraging that the average
mass retention rate required by torque-limited growth on galaxy evolution time
scales lies within the range of plausible values.  More detailed modeling could
include an explicit dependence of the mass retention rate on the accretion
mode and should account for the effects of AGN feedback on the host galaxy.

We have shown that there is no need for a strong redshift dependence of the
mass retention rate to regulate black hole growth, at least for the
range of black hole masses considered here.
This suggests that, to first order,
the net effect of winds and outflows from the accretion flow is to suppress
the instantaneous accretion rate by an average constant factor, contributing
significantly to the normalization of the AGN main sequence but only to
second order in the slope.  The rate of gas supply from galactic scales by
gravitational torques dominates the overall evolution of black hole growth on
cosmological timescales, sets the slope of the AGN main sequence, and yields
the average connection between SFR and AGN activity in star forming galaxies.

Assessing the full validity of our results will require explicit modeling of the interaction between accretion-driven outflows and the surrounding gas, which will be the subject of future work.
We speculate that AGN-driven outflows as required by torque-limited growth may have a limited impact on the host galaxy even if significant entrainment of cold ISM gas occurs.  
Jet heating of hot halo gas may however have a progressive long-term cumulative impact not only on the host galaxy but on black hole growth itself, linked to
the overall decrease of cosmological gas infall at lower redshifts and the
increasing frequency of radiatively inefficient accretion.  Cosmological
simulations have shown that preventing gas accretion in hot halos may yield
a galaxy red sequence and luminosity function as observed \citep{gabor12}.
Observations of powerful jets generated by a central black hole accreting at
low Eddington ratios imply heating rates that are comparable to the cooling
rates of hot gas in halos \citep{mcnamara07,fabian12}.  Radio-mode feedback
may thus prevent gas accretion into galaxies, reduce their overall star
formation, and possibly limit the amount of gas available for black hole
accretion \citep{okamoto08}.

The late time evolution of massive black holes in quiescent galaxies could, therefore, be
self-regulated by a ``true" large-scale feedback loop, where the gravitational torque model may no longer be an appropriate accretion parametrization.  Such a feedback loop in radio mode
is, however, unlikely to account for the majority of black hole growth,
expected to occur through radiatively efficient accretion \citep{sol82} in gas-rich star-forming galaxies \citep{heckman14}.
A hint for a switch in fuelling mechanism from torque-limited growth to
self-regulation in radio-mode may be given by observations of Eddington
ratio distributions at low redshift, with a characteristic transition from
log-normal to power-law distributions in star forming galaxies and passive
galaxies, respectively \citep{kauffmann09}.  We will explicitly address the
impact of AGN feedback on the co-evolution of black holes and galaxies in
future work, by performing self-consistent simulations of torque-limited
growth and AGN-driven outflows in a cosmological context.

\acknowledgments

We thank A. Escala, X. Fan, P. Hopkins, and B. Robertson for stimulating conversations.
We thank the anonymous referee for a thoughtful report that helped improve the paper.
F.\"O. gratefully acknowledges support from the Radcliffe Institute for
Advanced Study at Harvard University.  R.D. acknowledges support from the
South African Research Chair Initiative.  The simulations were run on the
University of Arizona's 512-processor SGI Altix system and the TACC Sun
Constellation Cluster (Ranger) at The University of Texas, Austin. This
work used the Extreme Science and Engineering Discovery Environment (XSEDE),
which is supported by National Science Foundation grant number OCI-1053575.
This work was supported by the National Science Foundation under grant numbers
AST-0907998 and AST-1108753 and through a grant from the Ahmanson Foundation.
This work was supported in part by the National Science Foundation under
Grant No. PHYS-1066293 and the hospitality of the Aspen Center for Physics.
This work was also supported by NASA Astrophysics Theory grants NNX12AH86G
and NNX10AJ95G.

\appendix

\section{Convergence of bulge--disk decomposition}\label{sec:app}

Different bulge--disk decomposition procedures are possible for the evaluation of black hole accretion rates through Equations~(\ref{eq:corr}) and~(\ref{eq:torque}) (Section~\ref{sec:accr}).
Here, we evaluate the numerical convergence properties of three such methods to show that the simple kinematic decomposition used in this work appears to be more robust at the resolution of our cosmological simulations relative to other commonly used methods.
For this, we make use of the eight re-simulated galaxies presented in \citet{ang13,ang14} for which two resolution levels are available, corresponding to 20 times and 2.5 times higher mass resolutions relative to our primary simulation in this work.  For each of these galaxies, we perform a bulge--disk decomposition at all available redshift snapshots (from early times down to $z = 2$), according to the following methods:
 
\begin{enumerate}

\item The bulge mass is calculated as double the mass of particles with $v_{\rm \phi} < 0$ within the effective radius of the galaxy, where $v_{\rm \phi}$ is the azimuthal velocity of each gas/star particle with respect to the rotation axis of the galaxy (defined as the direction of the total angular momentum within the effective radius).  This is the method employed in \citet{ang13} as well as in this work, formally identical to the bulge--disk decomposition procedure used by \citet{abadi03}.

\item The bulge component corresponds to the total mass of gas/star particles within the effective radius with orbital circularity parameter $\epsilon_{\rm J} \equiv J_{\rm z} / J_{\rm circ} \; < 0.8$, where $J_{\rm z}$ is the gas/star particle component of the specific angular momentum along the rotation axis of the galaxy and $J_{\rm circ} = r \times \sqrt{G\,M(<r)/r}$ is the specific angular momentum for a circular orbit at the particle radius $r$ corresponding to the enclosed mass $M(<r)$.  This method has been used in, e.g., \citet{governato09}, \citet{scannapieco09} and \citet{christensen14}.  Alternatively, we use a lower cut in the orbital circularity parameter to define the bulge component, $\epsilon_{\rm J} < 0.5$, similar to, e.g, \citet{tissera13} and \citet{pedrosa14}. 

\item We perform a standard two-component fit (S\'ersic bulge plus exponential disk) to the face-on azimuthally averaged mass surface density profiles of the stellar and gas components of simulated galaxies at all available redshift snapshots.  
The bulge mass is, then, computed as the integral of the S\'ersic profile within the effective radius of the galaxy.  Similar profile-fitting decompositions have been extensively applied to idealized galaxy simulations as well as high-resolution cosmological simulations \citep[e.g.,][]{robertson06a,hop09,christensen14}.

\end{enumerate}

\begin{figure}[t]
\begin{center}
\includegraphics[scale=0.5]{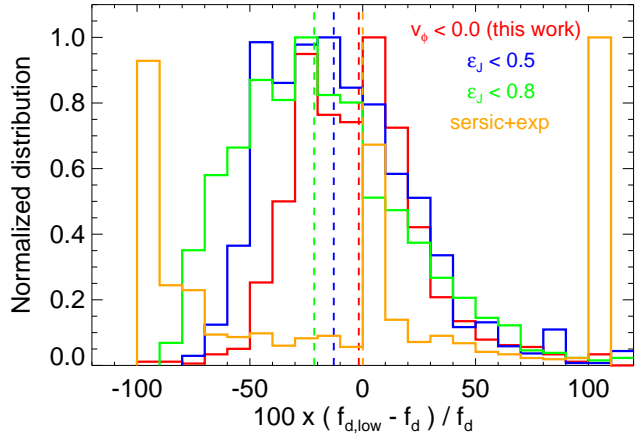}
\end{center}
\caption{Percentage difference between the disk fractions obtained for the low-resolution ($f_{\rm d,low}$) and high-resolution ($f_{\rm d}$) zoom-in simulations of \citet{ang13,ang14} according to different bulge--disk decomposition methods.  We calculate $\Delta_{\rm d} = 100 \times (f_{\rm d,low} - f_{\rm d}) / f_{\rm d}$ for each galaxy at each time step and plot the normalized distribution of $\Delta_{\rm d}$ values obtained for all galaxies in the redshift range $z = 2$--6.  Different criteria for the bulge--disk decomposition are considered, where the bulge mass is calculated as: (1) double the mass of particles with $v_{\rm \phi} < 0$ (this work; red), (2) the total mass of particles with $\epsilon_{\rm J} < 0.8$ (green) or  $\epsilon_{\rm J} < 0.5$ (blue), and (3) the mass contribution of the S\'ersic component from S\'ersic+exponential fits to the mass surface density profiles (orange).  
Vertical dashed lines of different colors indicate median percentage variations for each method.  
Note that upper and lower limits of $\pm \,100$\,\% have been imposed to the results from S\'ersic+exponential fitting.}
\label{fig:fig15}
\end{figure}

For all three methods, the disk fraction is calculated as $f_{\rm d} = 1 - M_{\rm bulge}(R_{\rm eff}) / M_{\rm tot}(R_{\rm eff})$, where $M_{\rm bulge}(R_{\rm eff})$ and $M_{\rm tot}(R_{\rm eff})$ correspond to the bulge mass and total mass for the gas and stellar components within the effective radius of the galaxy.  The resulting disk fractions are, then, compared between the high-resolution and low-resolution simulations of each galaxy as a function of redshift.  
Figure~\ref{fig:fig15} shows the normalized distribution of the percentage difference between the disk fractions obtained for the low- and high-resolution simulations according to the bulge--disk decomposition methods summarized above. 
As expected given the chaotic nature of hierarchical galaxy formation, we find a significant scatter in the distribution of disk fractions obtained from simulations of different resolution, which, nonetheless, produce galaxy morphologies in overall agreement.  
We find, however, a mild trend for lower disk fractions in the low-resolution simulations relative to the high-resolution simulations, with median percentage variations of --21\,\%, --13\,\%, and --2\,\% for the $\epsilon_{\rm J} < 0.8$, $\epsilon_{\rm J} < 0.5$, and $v_{\rm \phi} < 0$ bulge--disk decomposition conditions (methods 1 and 2).  That is, more strict conditions for the definition of the disk component in galaxies yield increasingly worse resolution convergence, with the simple kinematic decomposition used in this work ($v_{\rm \phi} < 0$) performing surprisingly well (at the expense of overestimating the disk fraction in the case of rotating bulges).  
Indeed, our modeling yields good numerical convergence relative to the black hole--galaxy scaling relations, as explicitly shown in \citet{ang13}.
Note that the two-component (S\'ersic plus exponential) profile fitting yields very inconsistent results between the disk fractions of high-resolution galaxies and their low-resolution analogs, owing to the degeneracy of fitting parameters, which cannot be appropriately constrained for a limited number of radial bins in low-resolution simulations.  While a more in-depth analysis of bulge--disk decomposition methods is beyond the scope of this work, this illustrates the challenge that increasingly complex models pose on the numerical robustness of cosmological simulations.   
The morphological decomposition procedure adopted in this work has the advantage of great simplicity together with good numerical convergence at the typical resolution of large scale cosmological simulations, making it a very attractive choice for on-the-fly calculations of black hole growth in galaxies across cosmic time.

%\clearpage

\end{document}